\documentclass[11pt]{econart}

\usepackage[utf8]{inputenc}
\usepackage{amsmath,amsthm,amssymb}
\usepackage{natbib}
\usepackage{graphicx}
\usepackage{setspace}
\usepackage{enumerate}
\usepackage{booktabs}
\usepackage{threeparttable}
\usepackage[table]{xcolor}
\usepackage{nicefrac}
\renewcommand*{\backref}[1]{}
\newcommand{\code}[1]{\texttt{#1}}
\newcommand{\nf}[2]{\nicefrac{#1}{#2}}

\DeclareMathOperator{\ARFIMA}{ARFIMA}

\definecolor{highlightgray}{gray}{0.92}
\newcommand{\largemse}[1]{\cellcolor{highlightgray}#1}

\begin{document}

\title{Semiparametric Estimation of Fractional Integration: An Evaluation of Local Whittle Methods}

\author{\uppercase{Jason R. Blevins} \\ The Ohio State University}

\pdfauthor{Jason R. Blevins}


\date{July 20, 2026}


\abstract{%
Fractionally integrated time series, exhibiting long memory with
slowly decaying autocorrelations, are frequently encountered in
economics, finance, and related fields.
Since the seminal work of \cite{robinson-1995-semiparametric},
a variety of semiparametric local Whittle estimators have been
proposed for estimating the memory parameter $d$, each with a
distinct range of validity and different robustness properties,
leaving applied researchers to decide which to use and
under what conditions.
This paper offers a practitioner's guide to six such estimators.
Using a common Monte Carlo design, we map how each estimator
behaves under short-run dynamics, unknown means,
and time trends---the conditions under which each remains reliable
and the characteristic way each breaks down.
This reveals a tension between efficiency and robustness: the exact
local Whittle estimator uniquely pairs the lowest asymptotic
variance with an unrestricted parameter range, but requires
the mean and trend to be handled with care.
We then illustrate these failure modes, along with the difficulties
introduced by structural breaks, on several macroeconomic,
financial, and climate time series, where a na\"{i}vely applied
estimator can report near-stationarity for a series that
better-matched methods identify as strongly nonstationary.
The resulting guidance on estimator choice and bandwidth selection
is anchored by exact reproductions of published results from this
literature, along with open source replication code and datasets.}

\maketitle

\section{Introduction}

Applied researchers who estimate the memory parameter $d$ of a
fractionally integrated series often encounter an uncomfortable
fact: different semiparametric estimators, applied to the same data,
can disagree sharply.
For French inflation (log-differenced CPI on an extended 1955--2025
sample), the six estimators we consider yield full-sample estimates $\hat{d}$ from
$0.12$ to $0.80$, a range that includes the nonstationarity boundary
at $0.50$ and leads to opposite conclusions about persistence.
For U.S. industrial production (log monthly index), the exact
local Whittle (ELW) estimator yields $\hat{d} = 0.10$, implying
behavior close to $I(0)$, while tapered estimators yield
$\hat{d} \approx 1.3$, implying strong nonstationarity.
Disagreements like these may reflect estimator limitations, features
of the data, or statistical uncertainty about $d$.
The goal of this paper is to evaluate these estimators and, in doing
so, to show practitioners how to identify such issues and how to
correct them.

The memory parameter is of interest in many fields such as
macroeconomics, finance, and climate econometrics.
A primary reason to estimate $d$, rather than simply difference or
detrend a series, is to preserve information for forecasting.
The hyperbolic decay of a long-memory process's autocovariances
carries predictive content at horizons where covariance-stationary
short-memory models do not \citep{bhardwaj-swanson-2006}.
The same consideration motivates the use of fractional differencing in
financial machine learning, to render a series stationary for predictive
modeling while preserving as much memory as possible
\citep{lopez-de-prado-2018}.

Since the seminal work of
\cite{granger-joyeux-1980} and \cite{hosking-1981}, who introduced
fractionally integrated processes to model intermediate persistence
between stationary and unit-root behavior, an extensive literature on
semiparametric estimation of $d$ has emerged.
For a broad survey of long memory and fractional integration in
econometrics, see \cite{baillie-1996}.
The early contributions
of \cite{geweke-porter-hudak-1983} and
\cite{robinson-1995-log-periodogram} developed log-periodogram
regressions for frequency-domain estimation of $d$. Alternatively,
by maximizing a quasi-likelihood function based on the Whittle
likelihood, the local Whittle (LW) estimator of \cite{kunsch-1987} and
\cite{robinson-1995-semiparametric} attains the conjectured\footnote{%
  \citet[p.~1640]{robinson-1995-semiparametric} conjectured that
  the semiparametric efficiency bound for this class of estimators
  is $\nf{1}{4}$.  Although no estimator has achieved
  lower asymptotic variance, the semiparametric efficiency bound
  has not been formally established.}
asymptotic variance lower bound of $\nf{1}{4}$ for stationary processes
with $d \in (-\nf{1}{2}, \nf{1}{2})$.

\cite{velasco-1999} and \cite{hurvich-chen-2000}
developed tapered LW variants that extended the method to
nonstationary processes, but at the cost of efficiency.
\cite{shimotsu-phillips-2005} introduced the ELW estimator,
which restored efficiency through exact fractional differencing
while extending the method to the entire parameter space.
The two-step ELW (2ELW) procedure of \cite{shimotsu-2010}
further extended the approach to processes with unknown means and
time trends.
The modified local Whittle estimator of \cite{hou-perron-2014} (LWLFC)
provided robustness to low-frequency contamination such as level shifts.

Other work has extended the scope of local Whittle methods along
several dimensions: to a wider memory range via a generalized discrete
Fourier transform \citep{abadir-distaso-giraitis-2007},
to reduced bias in the presence of short-run dynamics
\citep{robinson-henry-2003,andrews-sun-2004},
to multiple poles at seasonal and cyclical frequencies
\citep{arteche-2020,wingert-leschinski-sibbertsen-2020},
to perturbed fractional processes \citep{frederiksen-nielsen-nielsen-2012},
and to fractional cointegration \citep{nielsen-2007,shimotsu-2012}.

Previous authors have carried out finite-sample comparison studies
that provide a foundation for our work.
\cite{nielsen-frederiksen-2005} conducted an extensive Monte Carlo
comparison of parametric, semiparametric, and wavelet estimators.
Among bias-reducing semiparametric estimators, they found that
the local polynomial Whittle estimator of
\cite{andrews-sun-2004} and the bias-reduced log-periodogram
regression of \cite{andrews-guggenberger-2003} substantially
reduced bias in the presence of persistent short-run dynamics,
at the cost of increased variance.
A related study by \cite{garcia-enriquez-hualde-2019} compared
standard local Whittle estimation with the bias-reducing variants
of \cite{robinson-henry-2003} and \cite{andrews-sun-2004}.
They found that the benefits depend on the nature of the
underlying short-run dynamics.
These methods are helpful for moderate autoregressive short-run dynamics, but
they can underperform the simpler standard LW estimator when short-run
dynamics are mild.

Two features separate the present study from this earlier work.
First, neither assembles the full set of estimators compared here:
\cite{nielsen-frederiksen-2005} include a feasible exact local Whittle
estimator alongside the standard LW estimator, but the Hurvich--Chen-tapered
LW estimator of \cite{hurvich-chen-2000}, the two-step ELW estimator of
\cite{shimotsu-2010}, and the LWLFC estimator of
\cite{hou-perron-2014}---methods that accommodate unknown means,
trends, and level shifts---are included in neither comparison.
Second, neither extends the comparison beyond simulations into
empirical applications that connect the estimators to failures a
practitioner can diagnose in a specific series.

This paper addresses these gaps with a unified, practitioner-oriented
comparison of existing local Whittle estimators.
Our contribution is threefold.
First, we compare the finite-sample performance of six estimators
across an expanded parameter range: the standard LW, Velasco-tapered LW,
and ELW estimators, together with three methods not
previously included in such a comparison---the Hurvich--Chen-tapered LW,
two-step ELW, and LWLFC estimators.
To validate our implementations, we independently reproduce key
published Monte Carlo results for these estimators before extending
them in a unified design, with complete open-source replication
materials provided for every result.%
\footnote{All replication code and datasets are available at
  \url{https://github.com/jrblevin/lws}, and all estimates were
  computed using PyELW \citep{pyelw}, a Python library
  for local Whittle estimation.}

Second, we use the simulations to document the characteristic failure
modes of each estimator: the conditions---persistent short-run
dynamics, an unknown mean, a time trend, or nonstationarity beyond a
method's valid range---under which it becomes unreliable, and the
nature of its breakdown.
The ELW estimator illustrates this: it attains an asymptotic
variance one-third that of a higher-order tapered LW estimator,
but requires the mean to be handled correctly, and the ideal
correction depends on the memory parameter itself, a difficulty
that the two-step variant mitigates.

Third, through a series of empirical case studies, we show these
failure modes arising in practice and how to diagnose them.
For U.S. industrial production, the ELW objective function
produces a spurious global minimum near $d = 0.1$, while 2ELW's
adaptive mean correction reshapes the objective so that its
minimum is near the persistent value obtained by other estimators.
For French inflation, a series where genuine long memory and
structural breaks coexist, level shifts inflate the standard and
tapered estimates.
A unified comparison reveals patterns of agreement and disagreement
that often point to the data feature responsible.
We offer this as informal guidance rather than a formal procedure,
distilling from these cases practical lessons for estimator choice
given the features one has reason to suspect in the data.

The remainder of this paper is organized as follows.
Section~\ref{sec:methods} reviews the theoretical development of local
Whittle estimation methods and reports our replications of several main
results from the literature for each method.
Section~\ref{sec:mc} presents a new, cross-method Monte Carlo comparison
with controlled contaminations.
Section~\ref{sec:empirical} presents our extended empirical analyses based
on several macroeconomic, financial, and climate time series.
Section~\ref{sec:practical} summarizes our guidance on
selecting and applying local Whittle methods in practice.
Section~\ref{sec:conclusion} concludes.

\section{Local Whittle Estimators}
\label{sec:methods}

In this section, we briefly review six major local Whittle estimators.
We also independently reproduce key Monte Carlo results from
\cite{shimotsu-phillips-2005} and \cite{shimotsu-2010}.\footnote{%
  The Original and Replication columns in each panel of
  Tables~\ref{tab:mc:replications:lw}
  and~\ref{tab:mc:replications:elw} allow direct
  comparison and agree to within a few thousandths in most cells.
  The larger differences occur mainly at extreme parameters (e.g., $d = \pm 3.5$),
  far outside the estimators' valid ranges.}
Consider the fractionally integrated process $X_t$ defined as
\begin{equation}
  \label{eq:fi_process}
  (1-L)^{d_0} X_t = u_t \1\{t \geq 1\}, \quad t = 0, \pm 1, \pm 2, \dots,
\end{equation}
where $L$ is the lag operator,
$d_0 \in \R$ denotes the true memory parameter,
$\1\{\cdot\}$ denotes the indicator function, and
$u_t$ is a stationary, mean-zero process with spectral
density $f_u(\lambda)$.
Equation \eqref{eq:fi_process} is solved under the initialization
$X_t = 0$ for $t \leq 0$, which uniquely determines $X_t$ for
$t \geq 1$.\footnote{%
  The process defined in \eqref{eq:fi_process} is a ``Type~II''
  process, as opposed to a ``Type~I'' process, which applies the
  filter to a doubly-infinite sequence.}
The memory parameter $d_0$ determines the fundamental properties of $X_t$:
$I(0)$ behavior when $d_0 = 0$,
stationary long memory for $d_0 \in (0, \nf{1}{2})$,
nonstationary but mean-reverting dynamics for $d_0 \in [\nf{1}{2}, 1)$,
and the unit-root case when $d_0 = 1$.
Negative values of $d_0$ generate antipersistent processes,
while $d_0 > 1$ yields nonstationary processes that are not
mean-reverting.

The estimators we consider are frequency-domain methods that exploit
the distinctive spectral properties of long-memory processes
without specifying a parametric functional form for $f_u(\lambda)$.
The Whittle likelihood \citep{whittle-1951} provides a computationally
efficient approximation to the Gaussian likelihood by working in the
frequency domain, where the discrete Fourier coefficients are
asymptotically independent complex normal random variables
\citep[Theorem 4.4.1]{brillinger-1981}.
By focusing on low-frequency behavior, where the memory parameter
dominates the shape of the spectral density, local Whittle methods
offer two advantages: they avoid both the computationally intensive
matrix inversion required by time-domain maximum likelihood and the
need for a complete parametric specification.

\subsection{Local Whittle Estimation}
\label{sec:methods:lw}

The local Whittle estimator of \cite{robinson-1995-semiparametric} exploits
the power-law behavior of the spectral density near zero:
\begin{equation}
\label{eq:spectral_power_law}
f_X(\lambda) \sim G \lambda^{-2d_0} \quad \text{as} \quad \lambda \to 0^+,
\end{equation}
where
$a \sim b$ means that $\nf{a}{b} \to 1$ and
$G = f_u(0) > 0$.\footnote{For nonstationary processes
  with $d_0 \geq \nf{1}{2}$, the classical spectral density is
  undefined due to infinite variance.
  In this case, $f_X(\lambda)$ is interpreted as a pseudo-spectral
  density.}
This relationship captures the quintessential long-memory
properties while remaining agnostic about the specific
parametric form of the short-run dynamics.

The LW approach restricts attention to the first $m = m(n) \ll n$ Fourier frequencies
where \eqref{eq:spectral_power_law} holds most accurately, avoiding contamination
from short-run dynamics at higher frequencies.
We define the discrete Fourier transform of $a_t$ at frequencies
$\lambda_j = \nf{2\pi j}{n}$ as
\begin{equation*}
  w_a(\lambda_j) = \frac{1}{\sqrt{2\pi n}} \sum_{t=1}^n a_t e^{it\lambda_j}.
\end{equation*}
Let $I_a(\lambda_j) = |w_a(\lambda_j)|^2$ denote the periodogram.
The Whittle approximation to the (negative) Gaussian log-likelihood is
\begin{equation*}
\sum_{j=1}^m \left[ \log f_X(\lambda_j) + \frac{I_X(\lambda_j)}{f_X(\lambda_j)} \right].
\end{equation*}
Substituting the power-law approximation
$f_X(\lambda_j) \approx G\lambda_j^{-2d}$ for the unknown spectral density
and concentrating out the nuisance parameter $G$ yields
the local Whittle objective function:
\begin{equation}
\label{eq:lw_objective}
R(d) = \log\left(\frac{1}{m}\sum_{j=1}^m \lambda_j^{2d}I_X(\lambda_j)\right)
       - \frac{2d}{m}\sum_{j=1}^m \log\lambda_j.
\end{equation}

The local Whittle estimator is defined as:
\begin{equation*}
\hat{d}_{\mathrm{LW}} = \argmin_{d \in [\Delta_1, \Delta_2]} R(d),
\end{equation*}
where $[\Delta_1, \Delta_2]$ denotes the admissible parameter space.
Since the periodogram $I_X(\lambda_j)$ does not depend on $d$, it can
be precomputed from the data and so the objective function requires
only $O(m)$ operations per evaluation.
The objective function $R(d)$ is also convex in $d$
\citep[][Appendix A]{baum-hurn-lindsay-2020}, meaning that
any local minimum reached by an optimization
algorithm is also the global minimum.
 \cite{robinson-1995-semiparametric} established regularity conditions
under which for $d_0 \in (-\nf{1}{2}, \nf{1}{2})$,
the local Whittle estimator is asymptotically normal:
\begin{equation}
\label{eq:lw_asymptotics}
\sqrt{m}(\hat{d}_{\mathrm{LW}} - d_0) \dto \Normal\left(0, \nf{1}{4}\right).
\end{equation}
The asymptotic variance of $\nf{1}{4}$ implies an asymptotic standard error of
$\nf{1}{\sqrt{4m}}$ for $\hat{d}_{\mathrm{LW}}$.
\cite{velasco-1999} extended this analysis into the nonstationary region,
showing that the estimator remains consistent for
$d_0 \in (-\nf{1}{2}, 1)$ and asymptotically normal for
$d_0 \in (-\nf{1}{2}, \nf{3}{4})$.

The bandwidth parameter $m$ controls the bias-variance trade-off.
Increasing $m$ includes more frequencies which reduces variance, but
this increases bias since higher frequencies are influenced more by
short-run dynamics.
\cite{henry-robinson-1996} show that the theoretical
optimal rate is $m \propto n^{\nf{4}{5}}$, which minimizes
the mean squared error (MSE) of the estimator.
However, this rate can be sensitive to model
misspecification and short-run dynamics, leading practitioners
to often use more conservative choices
$m = \lfloor n^\alpha \rfloor$ with $\alpha < \nf{4}{5}$.
See \cite{henry-2001}, \cite{baillie-kapetanios-papailias-2014},
and particularly the discussion in \cite{arteche-orbe-2017} for
further developments on bandwidth selection, which we
revisit in Section~\ref{sec:empirical}.

\begin{table}[!tp]
\centering
\singlespacing
\begin{threeparttable}
\caption{Replication of Published Monte Carlo Results: LW and Tapered LW Estimators}
\label{tab:mc:replications:lw}
\footnotesize
\begin{tabular}{r@{\hspace{1.5em}}rrr@{\hspace{1.5em}}rrr}
\toprule
\multicolumn{1}{c}{} & \multicolumn{3}{c}{Original} & \multicolumn{3}{c}{Replication} \\
\cmidrule(lr){2-4} \cmidrule(lr){5-7}
$d$ & Bias & S.D. & MSE & Bias & S.D. & MSE \\
\midrule
\multicolumn{7}{c}{\textit{Panel A: Local Whittle (LW)}} \\
\midrule
$-1.3$ & $ 0.4109$ & $ 0.2170$ & $ 0.2160$ & $ 0.4114$ & $ 0.2182$ & $ 0.2169$ \\
$-0.7$ & $ 0.0353$ & $ 0.0885$ & $ 0.0091$ & $ 0.0373$ & $ 0.0882$ & $ 0.0092$ \\
$-0.3$ & $-0.0027$ & $ 0.0781$ & $ 0.0061$ & $-0.0008$ & $ 0.0780$ & $ 0.0061$ \\
$ 0.0$ & $-0.0075$ & $ 0.0781$ & $ 0.0062$ & $-0.0060$ & $ 0.0775$ & $ 0.0060$ \\
$ 0.3$ & $-0.0066$ & $ 0.0785$ & $ 0.0062$ & $-0.0060$ & $ 0.0778$ & $ 0.0061$ \\
$ 0.7$ & $ 0.0099$ & $ 0.0812$ & $ 0.0067$ & $ 0.0095$ & $ 0.0812$ & $ 0.0067$ \\
$ 1.3$ & $-0.2108$ & $ 0.0982$ & $ 0.0541$ & $-0.2106$ & $ 0.0986$ & $ 0.0541$ \\
\midrule
\multicolumn{7}{c}{\textit{Panel B: Velasco tapered LW, Bartlett taper}} \\
\midrule
$-3.5$ & $ 1.6126$ & $ 0.3380$ & $ 2.7148$ & $ 1.6231$ & $ 0.3326$ & $ 2.7450$ \\
$-2.3$ & $ 0.2155$ & $ 0.1726$ & $ 0.0762$ & $ 0.2173$ & $ 0.1734$ & $ 0.0773$ \\
$-1.7$ & $ 0.0259$ & $ 0.1235$ & $ 0.0159$ & $ 0.0279$ & $ 0.1211$ & $ 0.0154$ \\
$-1.3$ & $ 0.0081$ & $ 0.1211$ & $ 0.0147$ & $ 0.0090$ & $ 0.1203$ & $ 0.0146$ \\
$-0.7$ & $-0.0068$ & $ 0.1219$ & $ 0.0149$ & $-0.0048$ & $ 0.1200$ & $ 0.0144$ \\
$-0.3$ & $-0.0133$ & $ 0.1224$ & $ 0.0151$ & $-0.0097$ & $ 0.1201$ & $ 0.0145$ \\
$ 0.0$ & $-0.0138$ & $ 0.1224$ & $ 0.0152$ & $-0.0111$ & $ 0.1201$ & $ 0.0145$ \\
$ 0.3$ & $-0.0132$ & $ 0.1235$ & $ 0.0154$ & $-0.0103$ & $ 0.1201$ & $ 0.0145$ \\
$ 0.7$ & $-0.0068$ & $ 0.1227$ & $ 0.0151$ & $-0.0051$ & $ 0.1203$ & $ 0.0145$ \\
$ 1.3$ & $ 0.0140$ & $ 0.1232$ & $ 0.0154$ & $ 0.0170$ & $ 0.1222$ & $ 0.0152$ \\
$ 1.7$ & $ 0.0456$ & $ 0.1288$ & $ 0.0187$ & $ 0.0485$ & $ 0.1258$ & $ 0.0182$ \\
$ 2.3$ & $-0.1781$ & $ 0.1419$ & $ 0.0519$ & $-0.1774$ & $ 0.1410$ & $ 0.0514$ \\
$ 3.5$ & $-1.4541$ & $ 0.1338$ & $ 2.1322$ & $-1.4529$ & $ 0.1378$ & $ 2.1298$ \\
\midrule
\multicolumn{7}{c}{\textit{Panel C: Hurvich--Chen tapered LW (HC)}} \\
\midrule
$-3.5$ & $ 2.5889$ & $ 0.3037$ & $ 6.7946$ & $ 2.5920$ & $ 0.2988$ & $ 6.8075$ \\
$-2.3$ & $ 1.1100$ & $ 0.2893$ & $ 1.3157$ & $ 1.1104$ & $ 0.2882$ & $ 1.3160$ \\
$-1.7$ & $ 0.4474$ & $ 0.2154$ & $ 0.2466$ & $ 0.4493$ & $ 0.2149$ & $ 0.2481$ \\
$-1.3$ & $ 0.1551$ & $ 0.1231$ & $ 0.0392$ & $ 0.1547$ & $ 0.1239$ & $ 0.0393$ \\
$-0.7$ & $ 0.0278$ & $ 0.0957$ & $ 0.0099$ & $ 0.0294$ & $ 0.0959$ & $ 0.0101$ \\
$-0.3$ & $ 0.0100$ & $ 0.0971$ & $ 0.0095$ & $ 0.0134$ & $ 0.0972$ & $ 0.0096$ \\
$ 0.0$ & $ 0.0034$ & $ 0.0985$ & $ 0.0097$ & $ 0.0053$ & $ 0.0978$ & $ 0.0096$ \\
$ 0.3$ & $-0.0033$ & $ 0.1004$ & $ 0.0101$ & $-0.0007$ & $ 0.0983$ & $ 0.0097$ \\
$ 0.7$ & $-0.0066$ & $ 0.0994$ & $ 0.0099$ & $-0.0055$ & $ 0.0983$ & $ 0.0097$ \\
$ 1.3$ & $-0.0079$ & $ 0.0987$ & $ 0.0098$ & $-0.0050$ & $ 0.0974$ & $ 0.0095$ \\
$ 1.7$ & $ 0.0008$ & $ 0.0972$ & $ 0.0095$ & $ 0.0029$ & $ 0.0957$ & $ 0.0092$ \\
$ 2.3$ & $ 0.0528$ & $ 0.0981$ & $ 0.0124$ & $ 0.0557$ & $ 0.0981$ & $ 0.0127$ \\
$ 3.5$ & $-0.4079$ & $ 0.1142$ & $ 0.1795$ & $-0.4069$ & $ 0.1144$ & $ 0.1787$ \\
\bottomrule
\end{tabular}
\begin{tablenotes}
\footnotesize
\item Notes: Bias, standard deviation, and mean squared error of the estimates over 10,000 replications, each with $n = 500$ observations from $\ARFIMA(0,d,0)$ and $m = \lfloor n^{0.65} \rfloor = 56$ frequencies. ``Original'' columns report published values: Panel A from the right panel of Table 1 of Shimotsu and Phillips (2005), and Panels B and C from the right and left panels of their Table 2. ``Replication'' columns report results from our Python implementation.
\end{tablenotes}
\end{threeparttable}
\end{table}

Table~\ref{tab:mc:replications:lw} collects our replications of Monte
Carlo results, demonstrating the finite-sample properties of the standard
LW estimator and tapered variants (introduced below), with one panel per method.
These panels replicate Tables 1 and 2 of \cite{shimotsu-phillips-2005}.
Each sample consists of $n = 500$ simulated observations from an
$\ARFIMA(0,d,0)$ process.
The table shows the bias, standard deviation, and mean squared error
of the estimates over 10,000 replications.
The number of frequencies used was $m = \lfloor n^{0.65} \rfloor = 56$.
The ``Original'' columns report the published results while the
``Replication'' columns show the independent results from
our Python implementation.

Panel A contains the results for the LW estimator for a range of $d$ values.
Within the stationary and invertible range $d \in (-\nf{1}{2}, \nf{1}{2})$,
the estimator has very little bias,
standard deviations of about $0.078$ (moderately above the theoretical
asymptotic value of $\nf{1}{\sqrt{4m}} = 0.067$ due to the small bandwidth),
and low mean squared errors.
However, the theoretical limitations are apparent for the more extreme
values $d = -1.3$ and $d = 1.3$, where validity breaks down and
bias dominates the mean squared error.

\subsection{Extending the Range via Tapering}
\label{sec:methods:tapered}

The LW estimator achieves asymptotic normality over the range
$d_0 \in (-\nf{1}{2}, \nf{3}{4})$ with asymptotic variance
$\nf{1}{4}$; however, estimation outside of this limited range or in
the presence of deterministic components requires a different
approach.
To address this limitation, \cite{velasco-1999} (V) developed a class of
tapered LW estimators that trades efficiency for a wider validity range.

Formally, tapering involves weighting the original series $X_t$ by a
sequence $h_t$ that is symmetric around $\lfloor \nf{n}{2} \rfloor$,
with $\max_t h_t = 1$, and that decays smoothly to zero at the sample
boundaries.
The tapered periodogram
\begin{equation*}
I_h(\lambda_j) = |w_h(\lambda_j)|^2, \quad 
w_h(\lambda_j) = \frac{1}{\sqrt{2\pi n}} \sum_{t=1}^n h_t X_t e^{it\lambda_j},
\end{equation*}
replaces the standard periodogram $I_X(\lambda_j)$ in the objective function.
Because the periodogram treats the sample as repeating, a nonstationary
series creates an artificial jump between its two ends that contaminates low
frequencies (the very frequencies that identify $d_0$).
Tapering the ends of the series removes it.
For Velasco tapers, the periodogram uses subsampled frequencies
$\lambda_j = \nf{2\pi j}{n}$ with $j = p, 2p, 3p, \ldots, m$, where $p$ is the
taper order.
As with the standard LW estimator, evaluating these tapered objective functions
remains computationally inexpensive and the criterion functions are amenable to
simple scalar minimization algorithms.

Using tapers of higher orders $p$ extends the valid range of $d_0$ values
and provides robustness to trends of order $p-1$.
The triangular Bartlett window taper ($p=2$) is valid for
$d_0 \in (-\nf{1}{2}, \nf{3}{2})$ and robust to linear trends, while the
Zhurbenko--Kolmogorov (henceforth Kolmogorov) taper ($p=3$) is valid for
$d_0 \in (-\nf{1}{2}, \nf{5}{2})$ and robust to linear and quadratic time trends.
However, as mentioned, this robustness is at the expense of increased variance.
The variance of the limiting distribution of the tapered LW estimator
is $\nf{p \Phi_p}{4}$, where $\Phi_p$ is a constant that depends on
the order $p$ of the taper.\footnote{The limiting distribution and
  corresponding numerical constants are given in Theorem~6 of
  \cite{velasco-1999}.}
For $p=2$ with $\Phi_2=1.05000$, the asymptotic variance is $2 \times 1.05 = 2.1$ times
the LW variance;  for $p=3$ with $\Phi_3 = 1.00354$, the variance is $3.01$ times the LW variance.

Panel B of Table~\ref{tab:mc:replications:lw} replicates the right
panel of Table~2 from \cite{shimotsu-phillips-2005} to
illustrate the finite-sample behavior of the Bartlett-tapered LW estimator of
\cite{velasco-1999}.
Within the range $d \in [-1.7, 1.7]$ (which even includes values far
outside the method's valid range), the estimates have low bias and standard
deviations of around 0.120, roughly 25\% above the asymptotic value of
$\sqrt{\nf{2.1}{4 m}} = 0.097$.
As expected, at more extreme values of $d$ outside the method's
valid range, we see large biases and mean squared errors.

In Appendix~\ref{sec:appendix:tapers}, we compare the finite-sample
performance of all three tapers considered by \cite{velasco-1999}.
The comparison reflects the fundamental trade-off between bias
and variance due to tapering: the Bartlett taper achieves the
lowest variance, while the order-three Kolmogorov taper retains
low bias much deeper into the nonstationary range.
In the cross-method comparisons that follow, we use the Kolmogorov
taper for its wider range of validity.

\subsection{Efficiency Gains via Complex-Valued Tapering}
\label{sec:methods:hc}

\cite{hurvich-chen-2000} (HC) developed an alternative tapered local
Whittle estimator based on applying a complex-valued taper
$h_t = \tfrac{1}{2}\left(1 - e^{i 2\pi(t - 1/2)/n}\right)$, defined in
their equation~(3), to
first-differenced data $\Delta X_t = X_t - X_{t-1}$.
The HC-tapered LW estimator extends the valid range to
$d_0 \in (-\nf{1}{2}, \nf{3}{2})$ while achieving lower variance
than Velasco tapers.
It is also robust to linear trends, due to first-differencing the
data before applying the taper.
After estimating the memory parameter of the differenced data, the
researcher adds back one degree of integration to the resulting
estimate.
Because the tapered Fourier transform combines two adjacent untapered
ordinates, the HC method evaluates the spectrum at the shifted
frequencies $\lambda_{\tilde{j}} = 2\pi(j+0.5)/n$.
The estimator is defined as
\begin{equation*}
\hat{d}_{\mathrm{HC}} = 1 + \argmin_{d \in [\Delta_1 - 1, \Delta_2 - 1]} R_{\mathrm{HC}}(d),
\end{equation*}
where $R_{\mathrm{HC}}(d)$ has the same form as $R(d)$, but uses the
tapered periodogram of $\Delta X_t$.
The asymptotic variance of the HC estimator is $\nf{1.5}{4}$,
which is higher than $\nf{1}{4}$ for the LW estimator,
but lower than $\nf{2.1}{4}$ when using a second-order
Velasco taper.

In Panel C of Table~\ref{tab:mc:replications:lw}, we replicate the
HC-tapered LW estimator
results from \citet[Table 2]{shimotsu-phillips-2005}.
The HC estimator shows good finite-sample performance in its valid
range $d \in (-\nf{1}{2}, \nf{3}{2})$ and beyond, with degradation at
more extreme values.
We defer a full cross-method comparison until Section~\ref{sec:mc}.

\subsection{Exact Local Whittle Estimation}
\label{sec:methods:elw}

The standard local Whittle estimator assumes that the power-law
relationship between the periodograms of the observed and unobserved
processes---$I_u(\lambda_j) \approx \lambda_j^{2d_0} I_X(\lambda_j)$---holds
for frequencies near zero.
This approximation is valid when $\abs{d_0} < \nf{1}{2}$.
The exact local Whittle (ELW) estimator of \cite{shimotsu-phillips-2005}
instead works directly with fractionally differenced data,
avoiding this approximation.
From the model in \eqref{eq:fi_process}, we see that the periodogram
of the fractionally differenced series $\Delta^d X_t \equiv (1-L)^d X_t$
equals the periodogram of $u_t$ precisely when $d = d_0$, regardless of
the value of $d_0$.
By fractionally differencing the series for each candidate $d$ value
during optimization, the ELW estimator removes the approximation
error from the LW approach, extending consistency and asymptotic
normality to the full parameter space.

The fractional differencing operator is defined by the binomial expansion
\begin{equation*}
\Delta^d X_t = \sum_{k=0}^{t-1} \pi_k(d) X_{t-k},
\end{equation*}
where the coefficients $\pi_k(d)$ can be computed using
the recurrence
\begin{equation*}
\pi_0(d) = 1, \quad \pi_k(d) = \pi_{k-1}(d) \cdot \frac{k - 1 - d}{k}, \quad k \geq 1.
\end{equation*}
This operator transforms the data to remove $d$ orders of integration, and
when $d = d_0$, this operation recovers the short-memory process $u_t$
exactly.

The ELW objective function has a similar functional form as
\eqref{eq:lw_objective} for LW, but is based on
the periodogram $I_{\Delta^d X}(\lambda_j)$ of the fractionally differenced series:
\begin{equation}
\label{eq:elw_objective}
R_{\mathrm{ELW}}(d) = \log\left(\frac{1}{m}\sum_{j=1}^m I_{\Delta^d X}(\lambda_j)\right)
                   - \frac{2d}{m}\sum_{j=1}^m \log\lambda_j.
\end{equation}
Unlike for the LW objective, the periodogram
$I_{\Delta^d X}(\lambda_j)$ must be recomputed for each candidate $d$.
This introduces a nonlinear dependence on $d$ and, as a result,
the ELW objective is not guaranteed to be convex.

The primary advantage of the ELW estimator is that it remains valid
whether $d_0$ lies in the stationary region or far in the nonstationary
region, provided that the width of the optimization interval is
at most $\nf{9}{2}$.\footnote{%
  The width restriction arises because the proof of consistency
  requires different techniques when $|d - d_0| > \nf{1}{2}$, which
  have been established only for $|d - d_0| \leq \nf{9}{2}$.
  See the discussion on pp. 1894--1895 of \cite{shimotsu-phillips-2005}
  for details.}
Despite this, the ELW estimator retains the same asymptotic distribution
as the LW estimator:
\begin{equation*}
  \sqrt{m}(\hat{d}_{\mathrm{ELW}} - d_0) \dto \Normal\left(0, \nf{1}{4}\right).
\end{equation*}

The computational cost of ELW estimation is higher than for LW.
The latter only required one precomputed periodogram calculation and
no fractional differencing.
For the ELW estimator, for each candidate value of $d$ encountered during
optimization, one needs to recompute the fractional difference $\Delta^d X_t$ and
the corresponding periodogram $I_{\Delta^d X}$.
While computing $\Delta^d X_t$ na\"{i}vely requires $O(n^2)$ operations,
the fast fractional differencing algorithm
of \cite{jensen-nielsen-2014}, based on the fast Fourier transform,
reduces this to $O(n \log n)$.
The computational burden of ELW is therefore $O(n \log n)$ per objective
function evaluation, as compared to $O(m)$ for LW.
For this additional cost, we obtain a consistent estimator with asymptotic
variance $\nf{1}{4}$ across the full parameter space.

\begin{table}[t]
\centering
\singlespacing
\begin{threeparttable}
\caption{Replication of Published Monte Carlo Results: Exact Local Whittle Estimators}
\label{tab:mc:replications:elw}
\footnotesize
\begin{tabular}{r@{\hspace{1.5em}}rrr@{\hspace{1.5em}}rrr}
\toprule
\multicolumn{1}{c}{} & \multicolumn{3}{c}{Original} & \multicolumn{3}{c}{Replication} \\
\cmidrule(lr){2-4} \cmidrule(lr){5-7}
$d$ & Bias & S.D. & MSE & Bias & S.D. & MSE \\
\midrule
\multicolumn{7}{c}{\textit{Panel A: Exact local Whittle (ELW)}} \\
\midrule
$-3.5$ & $-0.0024$ & $ 0.0787$ & $ 0.0062$ & $-0.0014$ & $ 0.0777$ & $ 0.0060$ \\
$-2.3$ & $-0.0020$ & $ 0.0774$ & $ 0.0060$ & $-0.0015$ & $ 0.0777$ & $ 0.0060$ \\
$-1.7$ & $-0.0020$ & $ 0.0776$ & $ 0.0060$ & $-0.0015$ & $ 0.0778$ & $ 0.0061$ \\
$-1.3$ & $-0.0014$ & $ 0.0770$ & $ 0.0059$ & $-0.0015$ & $ 0.0778$ & $ 0.0061$ \\
$-0.7$ & $-0.0024$ & $ 0.0787$ & $ 0.0062$ & $-0.0015$ & $ 0.0778$ & $ 0.0061$ \\
$-0.3$ & $-0.0033$ & $ 0.0777$ & $ 0.0060$ & $-0.0015$ & $ 0.0779$ & $ 0.0061$ \\
$ 0.0$ & $-0.0029$ & $ 0.0784$ & $ 0.0061$ & $-0.0016$ & $ 0.0778$ & $ 0.0061$ \\
$ 0.3$ & $-0.0020$ & $ 0.0782$ & $ 0.0061$ & $-0.0015$ & $ 0.0779$ & $ 0.0061$ \\
$ 0.7$ & $-0.0017$ & $ 0.0777$ & $ 0.0060$ & $-0.0015$ & $ 0.0778$ & $ 0.0061$ \\
$ 1.3$ & $-0.0014$ & $ 0.0781$ & $ 0.0061$ & $-0.0014$ & $ 0.0779$ & $ 0.0061$ \\
$ 1.7$ & $-0.0025$ & $ 0.0780$ & $ 0.0061$ & $-0.0014$ & $ 0.0778$ & $ 0.0061$ \\
$ 2.3$ & $-0.0026$ & $ 0.0772$ & $ 0.0060$ & $-0.0012$ & $ 0.0779$ & $ 0.0061$ \\
$ 3.5$ & $-0.0016$ & $ 0.0770$ & $ 0.0059$ & $-0.0013$ & $ 0.0778$ & $ 0.0061$ \\
\midrule
\multicolumn{7}{c}{\textit{Panel B: Two-step exact local Whittle (2ELW)}} \\
\midrule
$ 0.0$ & $-0.0022$ & $ 0.0762$ & $ 0.0058$ & $-0.0019$ & $ 0.0758$ & $ 0.0057$ \\
$ 0.4$ & $ 0.0001$ & $ 0.0762$ & $ 0.0058$ & $ 0.0053$ & $ 0.0814$ & $ 0.0067$ \\
$ 0.8$ & $-0.0003$ & $ 0.0762$ & $ 0.0058$ & $-0.0012$ & $ 0.0764$ & $ 0.0058$ \\
$ 1.2$ & $-0.0006$ & $ 0.0755$ & $ 0.0057$ & $-0.0014$ & $ 0.0778$ & $ 0.0060$ \\
\bottomrule
\end{tabular}
\begin{tablenotes}
\footnotesize
\item Notes: Bias, standard deviation, and mean squared error of the estimates over 10,000 replications of $\ARFIMA(0,d,0)$. ``Original'' columns report published values; ``Replication'' columns report results from our Python implementation. Panel A replicates the left panel of Table 1 of Shimotsu and Phillips (2005), with $n = 500$ and $m = 56$. Panel B replicates Table 2 of Shimotsu (2010), with $n = 512$ and $m = 57$.
\end{tablenotes}
\end{threeparttable}
\end{table}

Panel A of Table~\ref{tab:mc:replications:elw} presents our
replication of the left panel of Table~1 from
\cite{shimotsu-phillips-2005}, demonstrating the finite-sample performance of
the ELW estimator over an extended range of $d$ values spanning
$d \in [-3.5, 3.5]$.
The results highlight the primary advantage of the ELW estimator:
under ideal conditions, it exhibits minimal bias, standard
deviations of about $0.078$,
and uniformly low mean squared errors for any value of $d$.

\subsection{Robustness to Unknown Mean and Trend}
\label{sec:methods:2elw}

The ELW estimator assumes that the stochastic process $X_t$ has zero
mean and does not contain a trend.
However, real economic time series may have nonzero means and time trends.
This led \cite{shimotsu-2010} to develop the two-step exact local Whittle
(2ELW) estimator, which addresses both issues while preserving the
$\Normal(0, \nf{1}{4})$ limiting distribution, although with a limited
parameter range.
The valid range depends on whether a trend is included or not,
as we will discuss below.

First, consider the model with an unknown mean:
\begin{equation*}
  X_t = \mu_0 + X_t^0, \quad X_t^0 = (1-L)^{-d_0} u_t \1\{t \geq 1\},
\end{equation*}
where $\mu_0$ is an unknown constant.
The problem this introduces is subtle.
The best way to handle the mean depends on the unknown value
of $d_0$ itself.
For stationary processes ($\abs{d_0} < \nf{1}{2}$), the sample
average $\hat{\mu} = \bar{X}$ is a reasonable estimator for $\mu_0$.
For nonstationary, highly persistent processes ($d_0 \geq \nf{1}{2}$),
the first observation $\hat{\mu} = X_1$ is a better estimator.
To address this indeterminacy, \cite{shimotsu-2010} proposed an
adaptive mean estimator $\tilde{\mu}(d)$ and a modified ELW objective
function based on fractionally differencing the mean-corrected data
$X_t - \tilde{\mu}(d)$.
The estimator $\tilde{\mu}(d)$ is based on a weight function that
varies smoothly with the value of $d$:
\begin{equation*}
\tilde{\mu}(d) = w(d)\bar{X} + (1-w(d))X_1,
\end{equation*}
where $w(d)$ is the weight function
\begin{equation}
\label{eq:2elw:weight}
w(d) = \begin{cases}
1 & \text{if } d \leq \frac{1}{2}, \\
\frac{1}{2}[1 + \cos(4\pi d)] & \text{if } \frac{1}{2} < d < \frac{3}{4}, \\
0 & \text{if } d \geq \frac{3}{4}.
\end{cases}
\end{equation}
The intuition behind this weighting is as follows: $w(d) = 1$
for stationary processes, so $\bar{X}$ receives full weight.
ELW with $\hat{\mu} = \bar{X}$ remains consistent for $d_0 < 1$, but
it loses asymptotic normality for $d_0 \geq \nf{3}{4}$.
Using $X_1$ instead extends asymptotic normality to $d_0 < 2$,
so $w(d) = 0$ for $d \geq \nf{3}{4}$, placing full weight on $X_1$.
The weight function incorporates a smooth cosine transition between
these two extremes.

Estimation of the mean in this way significantly complicates the asymptotic
theory, so \cite{shimotsu-2010} proposed a two-step approach based
on a $\sqrt{m}$-consistent first-step estimator.
The first step applies one of the tapered LW estimators discussed above
to obtain an initial estimate $\hat{d}_T$.
The second step applies a single Newton--Raphson step to the modified
ELW objective function starting from $\hat{d}_T$.
\cite{shimotsu-2010} noted that the Newton--Raphson step can become
numerically unstable when the Hessian $R''(\hat{d}_T)$ takes very small
values, resulting in extremely large updates, and recommended using
$\max\{R''(\hat{d}_T), 2\}$ instead of the actual Hessian.
The lower bound avoids extremely large steps leading to extreme estimates
$\hat{d}_{2\text{ELW}}$.

For data believed to contain a polynomial trend of order $k$, the
procedure first removes the trend via an OLS regression of $X_t$ on
$(1, t, t^2, \ldots, t^k)$, and then applies the above two-step
procedure to the residuals.
\cite{shimotsu-2010} shows that the 2ELW estimator is consistent with
detrending, but the valid parameter range is restricted to
$d_0 \in (-\nf{1}{2}, \nf{7}{4})$, compared to $d_0 \in (-\nf{1}{2}, 2)$
for the unknown mean case.
The procedure achieves the same $\Normal(0, \nf{1}{4})$ limiting
distribution as LW within these ranges while maintaining robustness
to both unknown means and polynomial trends.

Panel B of Table~\ref{tab:mc:replications:elw} presents our
replication of the 2ELW Monte Carlo results from Table~2 of
\cite{shimotsu-2010}, which used $n = 512$ observations and
$m = 57$ frequencies.
The 2ELW estimator maintains good finite-sample properties, with
minimal bias and performance that remains stable across the
stationary ($d = 0.0, 0.4$) and nonstationary ($d = 0.8, 1.2$) regions.
Here we only consider specifications for which there are no short-run
dynamics ($\rho = 0$), but in Section~\ref{sec:mc} we investigate the
robustness of 2ELW and other estimators in a more thorough set of
benchmarks including cases with AR(1) short-run dynamics ($\rho > 0$),
unknown means, and time trends.

\subsection{Robustness to Low-Frequency Contamination}
\label{sec:methods:lwlfc}

In practice, time series may be contaminated by level shifts that
distort the spectrum at low frequencies.
\cite{diebold-inoue-2001} and \cite{granger-hyung-2004} showed that
such contaminations can manifest as spurious long memory.
To address this, \cite{hou-perron-2014} developed a modified local Whittle
estimator designed to be robust to low-frequency contamination (LWLFC)
such as random level shifts, deterministic level shifts,
and certain types of normalized trends of the form $h(\nf{t}{n})$,
which remain bounded as $t$ grows.
However, LWLFC is only valid for stationary processes with
$d_0 \in [0,\nf{1}{2})$, and it remains sensitive to deterministic
linear trends (Section~\ref{sec:mc:time_trend}).%
\footnote{Because of its specialized role and restricted range, LWLFC
  does not appear in every comparison below.
  We include it in the comprehensive Monte Carlo study
  (Section~\ref{sec:mc:comprehensive}) and the structural-break application
  (Section~\ref{sec:empirical:hc:breaks}), but omit it where the
  general-purpose estimators are the focus.}

Specifically, let $Y_t$ denote the observed time series, which is assumed
to follow
\begin{equation*}
  Y_t = \mu_0 + X_t + L_t,
\end{equation*}
where $\mu_0$ is a constant, $X_t$ is a fractionally integrated process with
memory parameter $d_0$ as in \eqref{eq:fi_process}, and $L_t$ captures
low-frequency disturbances.

The estimator introduces an additional parameter $\theta \geq 0$,
representing a contamination ratio: the variance of the
low-frequency contamination $L_t$ relative to that of the
long-memory component $X_t$.
The modified objective function is
\begin{equation*}
J(d, \theta) = \log\left(\frac{1}{m}\sum_{j=1}^m \frac{I_Y(\lambda_j)}{g_j}\right)
              + \frac{1}{m}\sum_{j=1}^m \log(g_j),
\end{equation*}
where $g_j = \lambda_j^{-2d} + \theta \lambda_j^{-2}/n$.
The term $\theta \lambda_j^{-2}/n$ corrects the spectral density
for the contribution of the low-frequency contamination.
The LWLFC estimator is defined as
\begin{equation*}
  (\hat{d}_{\mathrm{LWLFC}}, \hat{\theta}_{\mathrm{LWLFC}}) = \argmin_{(d,\theta) \in [0,\nf{1}{2}) \times [0,\infty)} J(d,\theta).
\end{equation*}
\cite{hou-perron-2014} state regularity conditions for which the
LWLFC estimator is consistent and asymptotically normal with
\begin{equation*}
  \sqrt{m}(\hat{d}_{\mathrm{LWLFC}} - d_0) \dto \Normal\left(0, \nf{1}{4}\right).
\end{equation*}
The asymptotic variance is the same as that of the standard LW estimator,
meaning that accounting for low-frequency contamination does not reduce
efficiency, only range.
If there is no contamination, then
$\hat{\theta}_{\mathrm{LWLFC}} \pto 0$ and the LWLFC estimator is
asymptotically equivalent to the standard LW estimator.

\section{Monte Carlo Comparisons Across Methods}
\label{sec:mc}

In this section, we compare the finite-sample performance of the six
local Whittle estimators across several data-generating processes.
We consider the impact of short-run dynamics, unknown means, and time trends,
each under varying degrees of fractional integration.
Additionally, in Appendix~\ref{sec:appendix:robustness} we consider
robustness to the sample size, the form of the short-run dynamics,
and the innovation distribution.

Our Monte Carlo specifications follow conventions established
in the literature.
We use a sample size of $n = 500$ with Gaussian innovations
and bandwidth $m = \lfloor n^{0.65} \rfloor$.
These choices are consistent with the baseline specifications in
\cite{shimotsu-phillips-2005}, \cite{shimotsu-2010}, and
\cite{nielsen-frederiksen-2005}.
\cite{hurvich-chen-2000} also used $n = 500$ and Gaussian innovations,
but with bandwidth $m = \lfloor 0.25 n^{0.80} \rfloor$.
To focus on the performance of estimators without bandwidth variability,
for now we use a fixed bandwidth rule.
We will evaluate data-driven bandwidth choices in empirical applications
later in Section~\ref{sec:empirical}.

\subsection{Comprehensive Estimator Comparison with Short-Run Dynamics}
\label{sec:mc:comprehensive}

To establish a baseline for cross-estimator comparison,
we revisit the Monte Carlo design from Table I of \cite{hurvich-chen-2000}.
Their experiment used 500 replications of simulated $\ARFIMA(1,d,0)$
processes with AR(1) short-run dynamics (autoregressive parameter
$\rho$), while we use 10,000 replications for greater precision.
We use bandwidth $m = 56$ following the standard rule
$m = \lfloor n^{0.65} \rfloor$ mentioned above.

We extend their original study by comparing all six estimators:
LW, V = Velasco (Kolmogorov), HC = Hurvich--Chen, ELW, 2ELW, and LWLFC.
Furthermore, we evaluate these estimators over a wider range of
33 parameter combinations
\begin{equation*}
  (d, \rho) \in \{-2.2, -1.8, -1.2, -0.6, -0.3, 0.0, 0.3, 0.6, 1.2, 1.8, 2.2\} \times \{0.0, 0.5, 0.8\}.
\end{equation*}
This allows for a more thorough evaluation across ranges of fractional
integration from strongly persistent to antipersistent.
The results, reported in Table~\ref{tab:mc_comprehensive}, reveal
several patterns regarding the relative performance of estimators
under short-run AR(1) dynamics.
MSE values larger than $0.05$ are shaded in gray.\footnote{This
  threshold is arbitrary and was chosen as a simple way to
  highlight obvious failures.}

{\setlength{\tabcolsep}{3.5pt}\begin{table}[tp]
\centering
\begin{threeparttable}
\caption{Comprehensive Estimator Comparison}
\label{tab:mc_comprehensive}
\scriptsize
\begin{tabular}{cc|rrrrrr|rrrrrr}
\toprule
&  & \multicolumn{6}{c|}{Bias} & \multicolumn{6}{c}{MSE} \\
\cmidrule(lr){3-8} \cmidrule(lr){9-14}
$d$ & $\rho$ & LW & V & HC & ELW & 2ELW & LWLFC & LW & V & HC & ELW & 2ELW & LWLFC \\
\midrule
-2.2 & 0.0 &   1.505 &   0.072 &   0.992 &  -0.002 &   1.198 &   1.259 & \largemse{  2.359} &   0.032 & \largemse{  1.064} &   0.006 & \largemse{  1.505} & \largemse{  1.605} \\
-1.8 & 0.0 &   0.991 &   0.057 &   0.548 &  -0.002 &   0.753 &   0.824 & \largemse{  1.067} &   0.030 & \largemse{  0.356} &   0.006 & \largemse{  0.617} & \largemse{  0.685} \\
-1.2 & 0.0 &   0.318 &   0.039 &   0.111 &  -0.003 &   0.271 &   0.229 & \largemse{  0.138} &   0.028 &   0.024 &   0.006 & \largemse{  0.091} & \largemse{  0.057} \\
-0.6 & 0.0 &   0.020 &   0.031 &   0.025 &  -0.002 &   0.003 &  -0.007 &   0.007 &   0.028 &   0.010 &   0.006 &   0.006 &   0.008 \\
-0.3 & 0.0 &  -0.002 &   0.024 &   0.011 &  -0.002 &  -0.004 &  -0.024 &   0.006 &   0.027 &   0.010 &   0.006 &   0.006 &   0.009 \\
 0.0 & 0.0 &  -0.007 &   0.026 &   0.005 &  -0.002 &  -0.002 &  -0.033 &   0.006 &   0.027 &   0.010 &   0.006 &   0.006 &   0.010 \\
 0.3 & 0.0 &  -0.006 &   0.024 &  -0.002 &  -0.002 &  -0.001 &  -0.046 &   0.006 &   0.027 &   0.010 &   0.006 &   0.006 &   0.014 \\
 0.6 & 0.0 &   0.003 &   0.029 &  -0.007 &  -0.003 &   0.012 &  -0.082 &   0.006 &   0.027 &   0.010 &   0.006 &   0.006 &   0.045 \\
 1.2 & 0.0 &  -0.125 &   0.036 &  -0.008 &  -0.002 &  -0.002 &  -0.125 &   0.022 &   0.028 &   0.010 &   0.006 &   0.006 &   0.024 \\
 1.8 & 0.0 &  -0.738 &   0.063 &   0.004 &  -0.003 &  -0.003 &  -0.733 & \largemse{  0.560} &   0.031 &   0.009 &   0.006 &   0.006 & \largemse{  0.557} \\
 2.2 & 0.0 &  -1.160 &   0.095 &   0.038 &  -0.002 &  -0.002 &  -1.156 & \largemse{  1.358} &   0.036 &   0.011 &   0.006 &   0.006 & \largemse{  1.353} \\
\midrule
-2.2 & 0.5 &   1.318 &   0.191 &   0.881 &   0.100 &   1.088 &   1.206 & \largemse{  1.825} & \largemse{  0.063} & \largemse{  0.847} &   0.016 & \largemse{  1.248} & \largemse{  1.456} \\
-1.8 & 0.5 &   0.843 &   0.178 &   0.487 &   0.100 &   0.693 &   0.803 & \largemse{  0.778} & \largemse{  0.058} & \largemse{  0.275} &   0.016 & \largemse{  0.515} & \largemse{  0.645} \\
-1.2 & 0.5 &   0.275 &   0.161 &   0.181 &   0.101 &   0.275 &   0.237 & \largemse{  0.096} & \largemse{  0.052} &   0.043 &   0.016 & \largemse{  0.089} & \largemse{  0.060} \\
-0.6 & 0.5 &   0.110 &   0.150 &   0.132 &   0.100 &   0.100 &   0.099 &   0.019 & \largemse{  0.050} &   0.027 &   0.016 &   0.016 &   0.017 \\
-0.3 & 0.5 &   0.098 &   0.148 &   0.122 &   0.100 &   0.100 &   0.089 &   0.016 &   0.049 &   0.025 &   0.016 &   0.016 &   0.015 \\
 0.0 & 0.5 &   0.094 &   0.145 &   0.114 &   0.100 &   0.100 &   0.084 &   0.015 &   0.048 &   0.023 &   0.016 &   0.016 &   0.014 \\
 0.3 & 0.5 &   0.093 &   0.145 &   0.108 &   0.099 &   0.101 &   0.079 &   0.015 &   0.048 &   0.022 &   0.016 &   0.017 &   0.015 \\
 0.6 & 0.5 &   0.102 &   0.147 &   0.104 &   0.100 &   0.105 &   0.066 &   0.017 &   0.048 &   0.021 &   0.016 &   0.016 &   0.032 \\
 1.2 & 0.5 &  -0.096 &   0.158 &   0.102 &   0.099 &   0.099 &  -0.094 &   0.020 & \largemse{  0.052} &   0.021 &   0.016 &   0.016 &   0.020 \\
 1.8 & 0.5 &  -0.735 &   0.183 &   0.114 &   0.100 &   0.100 &  -0.732 & \largemse{  0.558} & \largemse{  0.060} &   0.023 &   0.016 &   0.016 & \largemse{  0.556} \\
 2.2 & 0.5 &  -1.160 &   0.207 &   0.144 &   0.100 &   0.100 &  -1.157 & \largemse{  1.358} & \largemse{  0.070} &   0.030 &   0.016 &   0.016 & \largemse{  1.354} \\
\midrule
-2.2 & 0.8 &   1.250 &   0.536 &   0.925 &   0.414 &   1.111 &   1.205 & \largemse{  1.638} & \largemse{  0.316} & \largemse{  0.905} & \largemse{  0.179} & \largemse{  1.278} & \largemse{  1.454} \\
-1.8 & 0.8 &   0.824 &   0.526 &   0.632 &   0.417 &   0.792 &   0.813 & \largemse{  0.723} & \largemse{  0.306} & \largemse{  0.418} & \largemse{  0.181} & \largemse{  0.649} & \largemse{  0.663} \\
-1.2 & 0.8 &   0.475 &   0.513 &   0.483 &   0.417 &   0.457 &   0.465 & \largemse{  0.234} & \largemse{  0.294} & \largemse{  0.243} & \largemse{  0.181} & \largemse{  0.215} & \largemse{  0.225} \\
-0.6 & 0.8 &   0.414 &   0.502 &   0.457 &   0.415 &   0.415 &   0.413 & \largemse{  0.179} & \largemse{  0.282} & \largemse{  0.220} & \largemse{  0.180} & \largemse{  0.179} & \largemse{  0.178} \\
-0.3 & 0.8 &   0.409 &   0.499 &   0.450 &   0.416 &   0.416 &   0.408 & \largemse{  0.174} & \largemse{  0.279} & \largemse{  0.214} & \largemse{  0.180} & \largemse{  0.180} & \largemse{  0.174} \\
 0.0 & 0.8 &   0.406 &   0.503 &   0.446 &   0.416 &   0.419 &   0.406 & \largemse{  0.172} & \largemse{  0.282} & \largemse{  0.211} & \largemse{  0.180} & \largemse{  0.184} & \largemse{  0.172} \\
 0.3 & 0.8 &   0.406 &   0.499 &   0.441 &   0.417 &   0.419 &   0.405 & \largemse{  0.172} & \largemse{  0.279} & \largemse{  0.206} & \largemse{  0.181} & \largemse{  0.183} & \largemse{  0.172} \\
 0.6 & 0.8 &   0.396 &   0.501 &   0.437 &   0.417 &   0.417 &   0.390 & \largemse{  0.164} & \largemse{  0.280} & \largemse{  0.202} & \largemse{  0.181} & \largemse{  0.181} & \largemse{  0.171} \\
 1.2 & 0.8 &  -0.035 &   0.511 &   0.435 &   0.417 &   0.417 &  -0.034 &   0.032 & \largemse{  0.289} & \largemse{  0.201} & \largemse{  0.181} & \largemse{  0.181} &   0.032 \\
 1.8 & 0.8 &  -0.731 &   0.526 &   0.441 &   0.416 &   0.416 &  -0.728 & \largemse{  0.558} & \largemse{  0.304} & \largemse{  0.205} & \largemse{  0.181} & \largemse{  0.180} & \largemse{  0.556} \\
 2.2 & 0.8 &  -1.158 &   0.551 &   0.462 &   0.416 &   0.416 &  -1.156 & \largemse{  1.355} & \largemse{  0.331} & \largemse{  0.224} & \largemse{  0.181} & \largemse{  0.180} & \largemse{  1.352} \\
\bottomrule
\end{tabular}
\begin{tablenotes}
\footnotesize
\item Notes: Monte Carlo results for $\ARFIMA(1,d,0)$ processes with $n=500$, $m=56$, 10,000 replications.
    Shaded cells indicate $\text{MSE} > 0.05$.
\item LW = Local Whittle, V = Velasco (Kolmogorov), HC = Hurvich--Chen, ELW = Exact Local Whittle, 2ELW = Two-step ELW, LWLFC = Local Whittle robust to low-frequency contamination.
\end{tablenotes}
\end{threeparttable}
\end{table}
}

Because we deliberately evaluated some estimators outside their valid
parameter ranges (e.g., 2ELW for $d < -0.5$), we can
characterize failure modes, which can be informative for practitioners
who do not know $d$ a priori.
In the absence of short-run dynamics (top panel, $\rho = 0$), the ELW estimator
performs best overall, with minimal bias
($-0.003$) and MSE values ($0.006$).
The 2ELW estimator has similar behavior, but fails with large bias
for antipersistent processes with $d < -0.6$.
This is consistent with its theoretical restriction to $d > -0.5$.
The standard LW estimator also fails beyond its valid range, with
bias of $-1.160$ and MSE of $1.358$ at $d = 2.2$, in line with its
known inconsistency for $d_0 > 1$ \citep{phillips-shimotsu-2004}.
Similarly, LWLFC shows good performance in its valid range $0 \leq d < 0.5$,
extending somewhat beyond, but fails outside that range.
The Velasco (V) estimator has positive bias larger than other estimators
in the valid range, while the Hurvich--Chen (HC) estimator shows
intermediate accuracy with generally smaller bias than V but larger
MSE than the exact methods.

The addition of moderate AR(1) persistence (middle panel, $\rho = 0.5$) changes the
relative performance landscape.
All estimators show upward bias, with the ELW and 2ELW methods
maintaining their advantage, although with higher bias ($\approx 0.100$).
The performance of the standard LW estimator remains good within its
valid range, and actually improves somewhat for highly
antipersistent cases, but still fails for extreme values of $d$,
with bias of $-0.735$ at $d = 1.8$ and $-1.160$ at $d = 2.2$.
The Velasco estimator maintains its stability but with higher
mean bias for all $d$ values.
The HC estimator again shows intermediate performance, with
bias increasing with $\abs{d}$ as before.

With stronger AR(1) persistence (bottom panel, $\rho = 0.8$), the performance of
all estimators suffers.
Bias for central $\abs{d}$ values exceeds 0.40 for most methods
and MSE values increase by an order of magnitude.
The ELW estimator has the smallest uniform bias ($\approx 0.416$),
while the Velasco estimator exhibits bias around 0.500 for most $d$ values.
The LW estimator's behavior is somewhat erratic, with essentially the
lowest bias of all estimators at $d=1.2$ but the highest bias of all
at $d = 2.2$ and $d = -2.2$.
As before, the performance of 2ELW is similar to ELW for central
values of $d$, but it fails more severely for large negative $d$.

In terms of overall robustness to AR(1) dynamics, the exact methods
(ELW and 2ELW) show the most stability across the parameter space.
Their performance degrades smoothly as autocorrelation increases.
Although the behavior of the Velasco estimator is also very
uniform across $d$ values, the bias and MSE are generally higher
than those of ELW and 2ELW, while the HC estimator shows intermediate
robustness and more sensitivity under antipersistence.
The standard LW and LWLFC estimators are most vulnerable to
autocorrelation when $\abs{d}$ is large, noting that such values
are beyond their valid range.
However, the exact methods' stability reflects their wider range of
validity rather than robustness to short-run dynamics:
within the stationary range they do not improve on the standard LW
estimator, whose bias at central $d$ is marginally smaller
(e.g., at $d = 0$ with $\rho = 0.8$, $0.406$ for LW against
$0.416$ for ELW).
Among these estimators,
we therefore recommend using exact local Whittle methods
whenever the likely range of $d$ is uncertain, with the
exception of 2ELW when strong antipersistence is possible.
Researchers may also consider the local polynomial Whittle estimator
of \cite{andrews-sun-2004}, which is designed to reduce this short-run
bias at the cost of higher variance, though it is restricted to
stationary processes with $\abs{d} < 0.5$.
Using a more conservative or data-driven bandwidth is another simple
step towards robustness.

Additional robustness simulations reported in
Appendix~\ref{sec:appendix:robustness} indicate that this ranking is
invariant to the short-run form (MA(1) and ARMA(1,1))---with the
magnitude of the bias being largely driven by proximity to an
autoregressive unit root---and to heavy-tailed and conditionally
heteroskedastic innovations.

\subsection{Sampling Distribution Comparison}

\begin{figure}[t!]
\centering
\includegraphics[width=\textwidth]{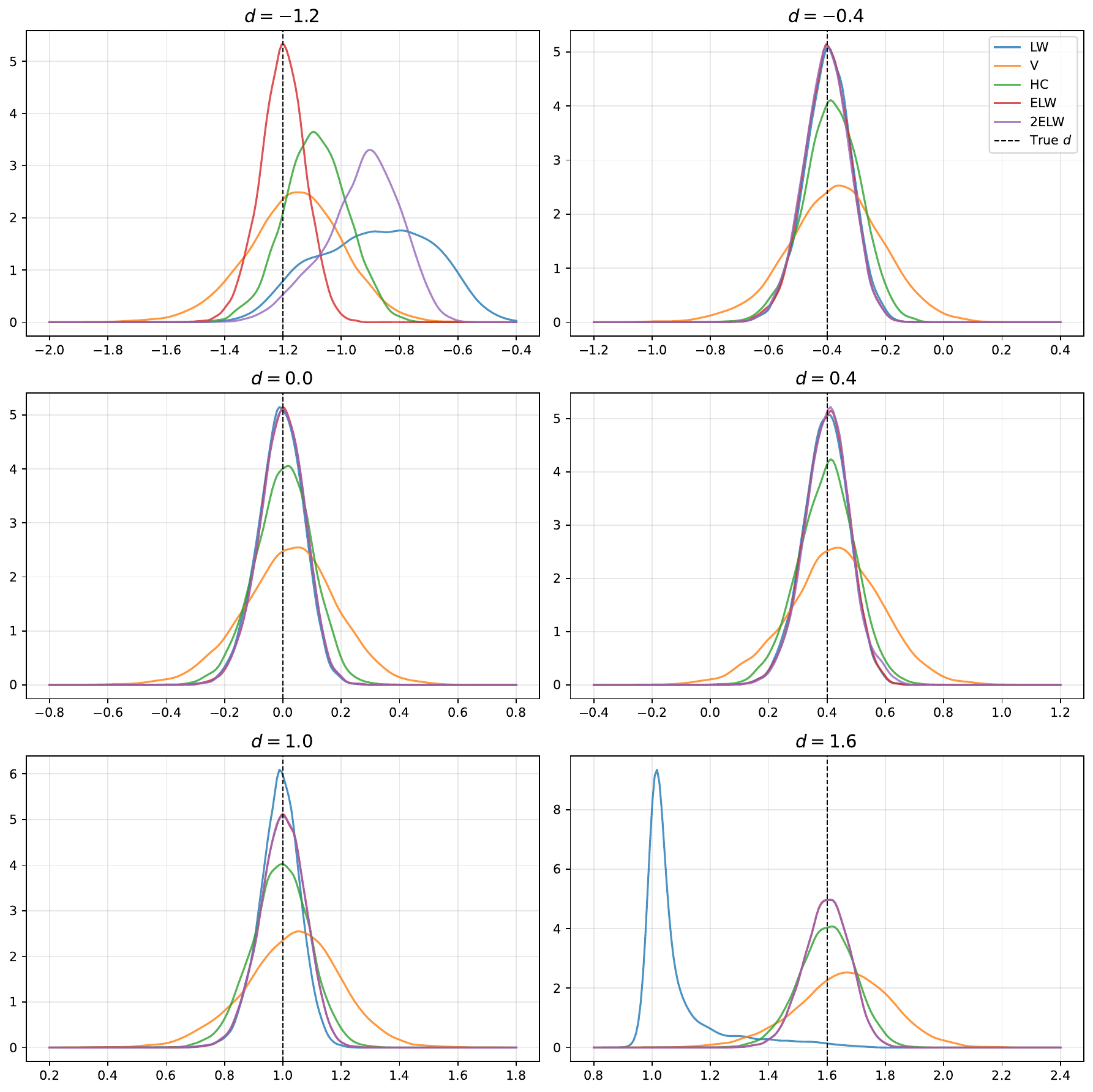}
\caption{Sampling distributions of local Whittle estimators with $n=500$ over 10,000 replications. V denotes the Velasco estimator with Kolmogorov taper.}
\label{fig:mc:distributions}
\end{figure}

Figure~\ref{fig:mc:distributions} presents kernel density estimates of the
sampling distributions for five estimators (LW, V, HC, ELW, and
2ELW) based on 10,000 replications of $\ARFIMA(0,d,0)$ processes with $n=500$
observations and bandwidth $m = \lfloor n^{0.65} \rfloor$.
The analysis spans six representative values of the memory parameter:
$d \in \{-1.2, -0.4, 0.0, 0.4, 1.0, 1.6\}$, covering antipersistent,
stationary long memory, and strongly persistent processes.

For the stationary models ($\abs{d} < 0.5$), in accordance with theory,
all estimators have approximately symmetric distributions centered
around the true parameter value.
In the case of strongly nonstationary processes ($d=1.6$), the LW
estimator is inconsistent and concentrates at $d=1.0$
\citep[][Theorem 3.2]{phillips-shimotsu-2004}, while the exact methods
(ELW and 2ELW) remain centered around the true values.
For the highly antipersistent case ($d=-1.2$), in accordance with the
theoretical properties, only the ELW estimator remains unbiased and
concentrated around the true value.
LW and 2ELW are biased upward, with means around $-0.9$.
The V and HC estimators also have positive bias, but to a lesser
degree.

\subsection{Robustness to Unknown Mean}
\label{sec:mc:unknown_mean}

{\setlength{\tabcolsep}{3.5pt}\begin{table}[tbp]
\centering
\begin{threeparttable}
\caption{Robustness to Unknown Mean}
\label{tab:mc_unknown_mean}
\footnotesize
\begin{tabular}{c|rrrrrr|rrrrrr}
\toprule
 & \multicolumn{6}{c|}{Baseline MSE ($\mu = 0$)} & \multicolumn{6}{c}{MSE Ratio ($\mu = 5$ / $\mu = 0$)} \\
\cmidrule(lr){2-7} \cmidrule(lr){8-13}
$d$ & LW & V & HC & ELW & 2ELW & LWLFC & LW & V & HC & ELW & 2ELW & LWLFC \\
\midrule
\multicolumn{13}{c}{\textit{Mean correction: None}} \\
\midrule
-2.2 & \largemse{  2.359} &   0.032 & \largemse{  1.064} &   0.006 & \largemse{  1.505} & \largemse{  1.605} &   0.993 &   0.987 &   0.999 & \largemse{1410.208} &   0.994 &   1.001 \\
-1.8 & \largemse{  1.077} &   0.029 & \largemse{  0.355} &   0.006 & \largemse{  0.618} & \largemse{  0.684} &   0.993 &   1.024 &   0.988 & \largemse{1024.138} &   0.993 &   0.999 \\
-1.2 & \largemse{  0.139} &   0.028 &   0.024 &   0.006 & \largemse{  0.091} & \largemse{  0.057} &   0.970 &   1.013 &   0.997 & \largemse{537.797} &   0.983 &   0.988 \\
-0.6 &   0.007 &   0.028 &   0.010 &   0.006 &   0.006 &   0.008 &   0.966 &   0.976 &   0.981 & \largemse{137.055} &   0.964 &   0.980 \\
-0.3 &   0.006 &   0.027 &   0.010 &   0.006 &   0.006 &   0.009 &   1.010 &   0.993 &   1.004 & \largemse{ 32.781} &   1.004 &   0.985 \\
0.0 &   0.006 &   0.027 &   0.010 &   0.006 &   0.006 &   0.010 &   0.978 &   1.006 &   0.977 & \largemse{  5.359} &   0.979 &   0.970 \\
0.3 &   0.006 &   0.027 &   0.010 &   0.006 &   0.006 &   0.014 &   1.016 &   0.985 &   1.020 & \largemse{ 14.515} &   1.016 &   0.997 \\
0.6 &   0.007 &   0.028 &   0.010 &   0.006 &   0.006 &   0.047 &   0.973 &   0.963 &   0.981 & \largemse{  2.369} &   0.974 &   0.936 \\
1.2 &   0.022 &   0.027 &   0.010 &   0.006 &   0.006 &   0.025 &   1.007 &   1.009 &   0.997 &   1.019 &   1.006 &   1.014 \\
1.8 & \largemse{  0.557} &   0.031 &   0.009 &   0.006 &   0.006 & \largemse{  0.554} &   1.006 &   1.003 &   1.011 &   1.032 &   1.027 &   1.006 \\
2.2 & \largemse{  1.359} &   0.036 &   0.011 &   0.006 &   0.006 & \largemse{  1.355} &   1.001 &   0.993 &   0.990 &   1.015 &   1.010 &   1.001 \\
\midrule
\multicolumn{13}{c}{\textit{Mean correction: $\hat{\mu} = \bar{X}$}} \\
\midrule
-2.2 & \largemse{  2.359} &   0.032 & \largemse{  1.064} & \largemse{  2.099} & \largemse{  1.505} & \largemse{  1.605} &   0.993 &   0.987 &   0.999 &   0.995 &   0.994 &   1.001 \\
-1.8 & \largemse{  1.077} &   0.029 & \largemse{  0.355} & \largemse{  0.980} & \largemse{  0.618} & \largemse{  0.684} &   0.993 &   1.024 &   0.988 &   0.996 &   0.993 &   0.999 \\
-1.2 & \largemse{  0.139} &   0.028 &   0.024 & \largemse{  0.121} & \largemse{  0.091} & \largemse{  0.057} &   0.970 &   1.013 &   0.997 &   0.977 &   0.983 &   0.987 \\
-0.6 &   0.007 &   0.028 &   0.010 &   0.006 &   0.006 &   0.008 &   0.966 &   0.976 &   0.981 &   0.963 &   0.964 &   0.980 \\
-0.3 &   0.006 &   0.027 &   0.010 &   0.006 &   0.006 &   0.009 &   1.010 &   0.993 &   1.004 &   1.004 &   1.004 &   0.985 \\
0.0 &   0.006 &   0.027 &   0.010 &   0.006 &   0.006 &   0.010 &   0.978 &   1.006 &   0.977 &   0.979 &   0.979 &   0.970 \\
0.3 &   0.006 &   0.027 &   0.010 &   0.006 &   0.006 &   0.014 &   1.016 &   0.985 &   1.020 &   1.017 &   1.016 &   0.997 \\
0.6 &   0.007 &   0.028 &   0.010 &   0.006 &   0.006 &   0.047 &   0.973 &   0.963 &   0.981 &   0.968 &   0.974 &   0.933 \\
1.2 &   0.022 &   0.027 &   0.010 &   0.012 &   0.006 &   0.025 &   1.007 &   1.009 &   0.997 &   1.009 &   1.006 &   1.014 \\
1.8 & \largemse{  0.557} &   0.031 &   0.009 & \largemse{  0.425} &   0.006 & \largemse{  0.554} &   1.006 &   1.003 &   1.011 &   1.008 &   1.027 &   1.006 \\
2.2 & \largemse{  1.359} &   0.036 &   0.011 & \largemse{  1.120} &   0.006 & \largemse{  1.355} &   1.001 &   0.993 &   0.990 &   1.002 &   1.010 &   1.001 \\
\midrule
\multicolumn{13}{c}{\textit{Mean correction: $\hat{\mu} = X_1$}} \\
\midrule
-2.2 & \largemse{  2.359} &   0.032 & \largemse{  1.064} & \largemse{  4.836} & \largemse{  1.505} & \largemse{  1.605} &   0.993 &   0.987 &   0.999 &   1.001 &   0.994 &   1.001 \\
-1.8 & \largemse{  1.077} &   0.029 & \largemse{  0.355} & \largemse{  3.128} & \largemse{  0.618} & \largemse{  0.684} &   0.993 &   1.024 &   0.988 &   1.003 &   0.993 &   0.999 \\
-1.2 & \largemse{  0.139} &   0.028 &   0.024 & \largemse{  1.296} & \largemse{  0.091} & \largemse{  0.057} &   0.970 &   1.013 &   0.997 &   0.999 &   0.983 &   0.987 \\
-0.6 &   0.007 &   0.028 &   0.010 & \largemse{  0.278} &   0.006 &   0.008 &   0.966 &   0.976 &   0.981 &   1.003 &   0.964 &   0.980 \\
-0.3 &   0.006 &   0.027 &   0.010 & \largemse{  0.058} &   0.006 &   0.009 &   1.010 &   0.993 &   1.004 &   0.998 &   1.004 &   0.985 \\
0.0 &   0.006 &   0.027 &   0.010 &   0.003 &   0.006 &   0.010 &   0.978 &   1.006 &   0.977 &   0.994 &   0.979 &   0.970 \\
0.3 &   0.006 &   0.027 &   0.010 &   0.009 &   0.006 &   0.014 &   1.016 &   0.985 &   1.020 &   1.000 &   1.016 &   0.997 \\
0.6 &   0.007 &   0.028 &   0.010 &   0.006 &   0.006 &   0.047 &   0.973 &   0.963 &   0.981 &   0.965 &   0.974 &   0.932 \\
1.2 &   0.022 &   0.027 &   0.010 &   0.006 &   0.006 &   0.025 &   1.007 &   1.009 &   0.997 &   1.006 &   1.006 &   1.015 \\
1.8 & \largemse{  0.557} &   0.031 &   0.009 &   0.006 &   0.006 & \largemse{  0.554} &   1.006 &   1.003 &   1.011 &   1.026 &   1.027 &   1.005 \\
2.2 & \largemse{  1.359} &   0.036 &   0.011 &   0.006 &   0.006 & \largemse{  1.355} &   1.001 &   0.993 &   0.990 &   1.010 &   1.010 &   1.001 \\
\bottomrule
\end{tabular}
\begin{tablenotes}
\footnotesize
\item Notes: MSE results for $\ARFIMA(0,d,0)$ with $n=500$, $m=56$, 10,000 replications.
    Shaded cells indicate $\text{MSE} > 0.05$ or $\text{MSE Ratio} > 2.0$.
\item LW = Local Whittle, V = Velasco (Kolmogorov), HC = Hurvich--Chen, ELW = Exact Local Whittle, 2ELW = Two-step ELW with adaptive mean estimation applied to original series, LWLFC = Local Whittle robust to low-frequency contamination.
\end{tablenotes}
\end{threeparttable}
\end{table}
}

Next, we consider the sensitivity of the estimators to unknown population means.
We compare performance across three approaches to mean correction:
no correction ($\hat{\mu} = 0$),
sample mean correction ($\hat{\mu} = \bar{X}$),
and first-observation correction ($\hat{\mu} = X_1$).
Table~\ref{tab:mc_unknown_mean} shows baseline MSE values (with $\mu = 0$) and
MSE ratios (with $\mu = 5$ relative to $\mu = 0$) for simulated $\ARFIMA(0,d,0)$
processes with $d \in [-2.2, 2.2]$.
Ratios near unity indicate robustness to nonzero means;
larger values indicate performance degradation.
Since the mean is unknown a priori, for each correction we transform
the data as $\tilde{X}_t = X_t - \hat{\mu}$ before applying the estimator.
We apply this transformation in both cases $\mu = 0$ and $\mu = 5$.
The only exception is 2ELW, which receives the original data and applies
its own internal adaptive mean correction.

In the top panel, with no mean correction, the presence of a nonzero mean reveals
differences in robustness.
The Velasco estimator shows both good baseline performance (MSE $\approx 0.03$) and
robustness (MSE ratios near unity) for all $d$ values.
The LW, HC, 2ELW, and LWLFC estimators also exhibit MSE ratios near
unity, but their baseline MSE values are elevated at extreme $d$ values
(e.g., LW has baseline MSE of 2.359 at $d = -2.2$), where a ratio near
unity indicates consistently poor performance rather than robustness.
In contrast, ELW has extremely low baseline MSE values
($\approx 0.006$ across $d$), but exhibits severe sensitivity to
mean contamination for $d \leq 0.6$, especially for antipersistent processes
(MSE ratio of 5.4 at $d = 0.0$, increasing to 1410 at $d = -2.2$),
with robustness only for $d \geq 1.2$.

In the middle panel, we apply a sample mean correction ($\hat{\mu} = \bar{X}$).
All estimators have MSE ratios near unity, meaning that
their performance is similar whether the true mean is zero or nonzero.
However, the sample mean correction itself (unnecessarily demeaning data with
$\mu = 0$) degrades the baseline performance of
ELW, with MSE of 2.099 at $d = -2.2$ and 1.120 at $d = 2.2$, while the
other estimators maintain their baseline performance levels.

In the bottom panel, we consider the effects of the first-observation
correction ($\hat{\mu} = X_1$).
Similar to the case of the sample mean correction, all estimators have
MSE ratios near unity.
However, for ELW the baseline MSE effects are asymmetric:
baseline MSE is elevated for $d < 0$ (4.836 at $d = -2.2$), but
remains at the low level of 0.006 for $d \geq 0$.

Overall, these results demonstrate that the Velasco-tapered LW estimator
provides remarkable robustness to unknown means, corrected or
not, over the full range of $d$ values.
LW, HC, and 2ELW are largely unaffected, but have ranges of $d$ values
where they perform much worse than Velasco.
For ELW, the best mean correction critically depends on the expected
range of $d$, which is precisely the intuition behind the 2ELW
estimator's adaptive mean correction.

\subsection{Robustness to Linear Time Trends}
\label{sec:mc:time_trend}

{\setlength{\tabcolsep}{3.5pt}\begin{table}[t!]
\centering
\begin{threeparttable}
\caption{Robustness to Time Trend}
\label{tab:mc_time_trend}
\footnotesize
\begin{tabular}{c|rrrrrr|rrrrrr}
\toprule
& \multicolumn{6}{c|}{Baseline MSE ($\beta = 0.0$)} & \multicolumn{6}{c}{MSE Ratio ($\beta = 0.05$ / $\beta = 0.0$)} \\
\cmidrule(lr){2-7} \cmidrule(lr){8-13}
$d$ & LW & V & HC & ELW & 2ELW & LWLFC & LW & V & HC & ELW & 2ELW & LWLFC \\
\midrule
\multicolumn{13}{c}{\textit{Trend correction: None}} \\
\midrule
-2.2 & \largemse{  2.359} &   0.032 & \largemse{  1.064} &   0.006 & \largemse{  1.530} & \largemse{  1.605} & \largemse{  4.311} &   1.019 &   0.999 & \largemse{1487.482} &   0.995 & \largemse{  6.333} \\
-1.8 & \largemse{  1.077} &   0.029 & \largemse{  0.355} &   0.006 & \largemse{  0.626} & \largemse{  0.684} & \largemse{  7.213} &   1.027 &   0.988 & \largemse{1099.768} &   0.993 & \largemse{ 11.358} \\
-1.2 & \largemse{  0.139} &   0.028 &   0.024 &   0.006 & \largemse{  0.102} & \largemse{  0.057} & \largemse{ 34.315} &   1.013 &   0.997 & \largemse{641.153} &   0.987 & \largemse{ 83.832} \\
-0.6 &   0.007 &   0.028 &   0.010 &   0.006 &   0.007 &   0.008 & \largemse{336.676} &   0.976 &   0.981 & \largemse{287.097} &   0.973 & \largemse{293.749} \\
-0.3 &   0.006 &   0.027 &   0.010 &   0.006 &   0.006 &   0.009 & \largemse{259.778} &   0.993 &   1.004 & \largemse{170.877} &   0.995 & \largemse{181.603} \\
0.0 &   0.006 &   0.027 &   0.010 &   0.006 &   0.007 &   0.010 & \largemse{147.260} &   1.006 &   0.977 & \largemse{ 80.003} &   0.974 & \largemse{ 87.692} \\
0.3 &   0.006 &   0.027 &   0.010 &   0.006 &   0.007 &   0.014 & \largemse{ 68.177} &   0.985 &   1.020 & \largemse{ 25.295} &   1.012 & \largemse{ 29.449} \\
0.6 &   0.007 &   0.028 &   0.010 &   0.006 &   0.007 &   0.047 & \largemse{ 16.721} &   0.963 &   0.981 & \largemse{  2.257} &   0.975 & \largemse{  3.524} \\
1.2 &   0.022 &   0.027 &   0.010 &   0.006 &   0.006 &   0.025 &   1.044 &   1.009 &   0.997 &   1.010 &   1.003 &   1.072 \\
1.8 & \largemse{  0.557} &   0.031 &   0.009 &   0.006 &   0.006 & \largemse{  0.554} &   1.006 &   1.003 &   1.011 &   1.026 &   0.997 &   1.006 \\
2.2 & \largemse{  1.359} &   0.036 &   0.011 &   0.006 &   0.011 & \largemse{  1.355} &   1.001 &   0.993 &   0.990 &   1.010 &   0.998 &   1.001 \\
\midrule
\multicolumn{13}{c}{\textit{Trend correction: linear OLS detrending}} \\
\midrule
-2.2 & \largemse{  2.551} &   0.032 & \largemse{  1.064} & \largemse{  2.443} & \largemse{  1.530} & \largemse{  1.585} &   0.994 &   0.987 &   0.999 &   0.994 &   0.995 &   1.001 \\
-1.8 & \largemse{  1.211} &   0.029 & \largemse{  0.355} & \largemse{  1.174} & \largemse{  0.626} & \largemse{  0.677} &   0.994 &   1.024 &   0.988 &   0.994 &   0.993 &   1.000 \\
-1.2 & \largemse{  0.180} &   0.028 &   0.024 & \largemse{  0.177} & \largemse{  0.102} & \largemse{  0.054} &   0.973 &   1.013 &   0.997 &   0.974 &   0.987 &   0.988 \\
-0.6 &   0.008 &   0.028 &   0.010 &   0.007 &   0.007 &   0.009 &   0.977 &   0.976 &   0.981 &   0.973 &   0.973 &   0.990 \\
-0.3 &   0.006 &   0.027 &   0.010 &   0.006 &   0.006 &   0.009 &   1.001 &   0.993 &   1.004 &   0.995 &   0.995 &   0.985 \\
0.0 &   0.007 &   0.027 &   0.010 &   0.007 &   0.007 &   0.010 &   0.974 &   1.006 &   0.977 &   0.974 &   0.974 &   0.969 \\
0.3 &   0.007 &   0.027 &   0.010 &   0.007 &   0.007 &   0.013 &   1.012 &   0.985 &   1.020 &   1.012 &   1.012 &   0.987 \\
0.6 &   0.008 &   0.028 &   0.010 &   0.007 &   0.007 &   0.029 &   0.974 &   0.963 &   0.981 &   0.978 &   0.975 &   0.952 \\
1.2 &   0.015 &   0.027 &   0.010 &   0.009 &   0.006 &   0.020 &   0.989 &   1.009 &   0.997 &   0.980 &   1.003 &   1.043 \\
1.8 & \largemse{  0.317} &   0.031 &   0.009 & \largemse{  0.317} &   0.006 & \largemse{  0.308} &   1.012 &   1.003 &   1.011 &   1.013 &   0.997 &   1.013 \\
2.2 & \largemse{  0.870} &   0.036 &   0.011 & \largemse{  1.029} &   0.011 & \largemse{  0.837} &   1.011 &   0.993 &   0.990 &   1.006 &   0.998 &   1.009 \\
\bottomrule
\end{tabular}
\begin{tablenotes}
\footnotesize
\item Notes: MSE results for $\ARFIMA(0,d,0)$, $n=500$, $m=56$, 10,000 replications.
    Shaded cells indicate $\text{MSE} > 0.05$ or $\text{MSE Ratio} > 2.0$.
\item LW = Local Whittle, V = Velasco (Kolmogorov), HC = Hurvich--Chen, ELW = Exact Local Whittle, 2ELW = Two-step ELW with linear detrending and adaptive mean estimation applied to original series, LWLFC = Local Whittle robust to low-frequency contamination.
\end{tablenotes}
\end{threeparttable}
\end{table}
}

The Monte Carlo results in Table~\ref{tab:mc_time_trend} examine the sensitivity
of the estimators to linear deterministic time trends across the parameter range.
The experiment involves simulated $\ARFIMA(0,d,0)$ processes under two scenarios:
no trend ($\beta = 0.0$) and linear trend ($\beta = 0.05$).
For both specifications, we report the performance both with and without
detrending.
As in the previous section, this table displays baseline MSE values
(for $\beta = 0.0$) and MSE ratios ($\beta = 0.05$ relative to $\beta = 0.0$).
For LW, V, HC, and ELW, in the ``linear OLS detrending'' panel we apply the
estimators to preprocessed data that have been detrended (i.e., residuals from an
OLS regression on a constant and $t$).
In contrast, we always apply 2ELW to the original unprocessed data and allow it
to apply its own internal linear detrending.
As a result, the 2ELW results are identical in the top and bottom panels.

With no trend correction (top panel), the Velasco- and Hurvich--Chen-tapered
LW estimators demonstrate both good baseline MSE within their valid range and
robustness to a linear trend (i.e., MSE ratios are near unity).
In contrast, LW, ELW, and LWLFC are highly sensitive to the trend for
$d \leq 0.6$: MSE ratios for LW exceed 100 for $-0.6 \leq d \leq 0.0$,
and those for ELW climb as $d$ falls, reaching 1487 at $d = -2.2$.
The baseline MSE of 2ELW is elevated for $d \leq -1.2$, yet it shows good robustness to
the trend with MSE ratios near unity for all $d$ values.

Detrending by OLS (bottom panel) equalizes performance between trend scenarios,
with all estimators exhibiting MSE ratios near unity.
However, like demeaning, detrending when no trend is actually present
can be detrimental to baseline performance for ELW in particular:
baseline MSE increases from 0.006 to 2.443 at $d = -2.2$ and 1.029 at $d = 2.2$.

These results illustrate an important practical point:
the Velasco- and Hurvich--Chen-tapered LW estimators provide robustness to
linear deterministic trends without any preprocessing.
The internal trend correction of 2ELW is also effective.
On the other hand, the LW, ELW, and LWLFC estimators fail unless the
researcher detrends the data, and may still fail in spite of the correction
for more extreme values of $d$.

\section{Empirical Applications}
\label{sec:empirical}

We now turn from controlled Monte Carlo experiments to real economic data,
where the memory parameter is unknown and the data may violate ideal
assumptions.
This section replicates and extends the empirical analysis of
\cite{hurvich-chen-2000}.
Using their original datasets, but now comparing across multiple estimators,
we examine anomalous estimates, bandwidth selection, and
structural breaks.
Our emphasis is methodological: each series illustrates how the
estimators behave on real data and how to approach model
specification and estimator choice, with lessons that extend
beyond the particular series examined.

\subsection{Replication and Extension of Table III}
\label{sec:empirical:hc:replication}

\begin{table}[t!]
\centering
\begin{threeparttable}
\caption{Hurvich and Chen (2000) Datasets: LW Estimator Comparison}
\label{tab:emp_hurvich_chen}
\scriptsize
\begin{tabular}{lrr|c|ccccc}
\toprule
& & & Original & \multicolumn{5}{c}{Replication} \\
\cmidrule(lr){4-4} \cmidrule(lr){5-9}
Series & $n$ & $m$ & HC & LW & V & HC & ELW & 2ELW \\
\midrule
Global temp. & 1632 & 130 & $0.45$ & $0.50$ & $0.44$ & $0.45$ & $0.50$ & $0.47$ \\
& & & $(0.060)$ & $(0.044)$ & $(0.076)$ & $(0.060)$ & $(0.044)$ & $(0.044)$ \\
S\&P 500 & 8432 & 1383 & $0.99$ & $0.98$ & $0.98$ & $0.97$ & $0.99$ & $0.98$ \\
& & & $(0.018)$ & $(0.013)$ & $(0.023)$ & $(0.017)$ & $(0.013)$ & $(0.013)$ \\
Inflation, US & 491 & 40 & $0.57$ & $0.63$ & $0.66$ & $0.56$ & $0.64$ & $0.64$ \\
& & & $(0.123)$ & $(0.079)$ & $(0.137)$ & $(0.121)$ & $(0.079)$ & $(0.079)$ \\
Inflation, UK & 491 & 40 & $0.33$ & $0.46$ & $0.49$ & $0.34$ & $0.47$ & $0.47$ \\
& & & $(0.123)$ & $(0.079)$ & $(0.137)$ & $(0.121)$ & $(0.079)$ & $(0.079)$ \\
Inflation, FR & 491 & 40 & $0.67$ & $0.46$ & $0.82$ & $0.68$ & $0.50$ & $0.45$ \\
& & & $(0.123)$ & $(0.079)$ & $(0.137)$ & $(0.121)$ & $(0.079)$ & $(0.079)$ \\
Real wages, US & 492 & 35 & $1.43$ & $1.06$ & $1.56$ & $1.43$ & $0.06$ & $1.29$ \\
& & & $(0.121)$ & $(0.085)$ & $(0.147)$ & $(0.132)$ & $(0.085)$ & $(0.085)$ \\
Ind. prod., US & 492 & 100 & $1.34$ & $1.00$ & $1.31$ & $1.36$ & $0.10$ & $1.31$ \\
& & & $(0.075)$ & $(0.050)$ & $(0.087)$ & $(0.071)$ & $(0.050)$ & $(0.050)$ \\
\bottomrule
\end{tabular}
\begin{tablenotes}
\footnotesize
\item Notes: Estimates of $d$ with asymptotic standard errors in parentheses. The Original column shows results from Hurvich and Chen (2000) Table III for HC. Replication columns report results from multiple estimators: LW = Local Whittle, V = Velasco (Kolmogorov), HC = Hurvich--Chen, ELW = Exact Local Whittle, 2ELW = Two-step ELW. The replication HC column reports standard errors based on their variance approximation (rather than asymptotic standard errors) for direct comparison with the original.
\end{tablenotes}
\end{threeparttable}
\end{table}

We begin by replicating and extending the empirical analysis in Table III of
\cite{hurvich-chen-2000}, who applied their complex tapered LW estimator to
seven datasets including economic, financial, and environmental time series.
Motivated by their work, in Table~\ref{tab:emp_hurvich_chen} we compare
five LW estimators using the same datasets.\footnote{The reported
  number of observations is increased by one because \cite{hurvich-chen-2000}
  report the number of observations after first differencing.}

The global temperature series contains seasonally adjusted monthly temperatures
for the northern hemisphere from 1854--1989, computed as deviations from monthly
averages relative to 1950--1979 \citep{beran-1994}.
The S\&P 500 series consists of the natural logarithms of the daily stock index
levels from July 1962 to December 1995.
The remaining series were drawn from the International Monetary Fund's
International Financial Statistics database and consist of monthly
economic indicators from January 1957 to December 1997.
These include CPI-based inflation rates for the U.S., U.K., and France
(differences in logarithms of CPI), the logarithm of U.S. real manufacturing
wages,\footnote{We note one important exception in our replication.
  Since the original IMF data on U.S. real manufacturing wages no longer
  appear to be available, we instead use FRED average hourly earnings of
  production and nonsupervisory employees in manufacturing (CES3000000008)
  divided by CPI, which matches the published HC estimate.}
and the logarithm of U.S. industrial production.
All of these economic series were seasonally adjusted.
We intentionally maintain the original sample periods to permit
direct comparison with the published estimates.
In Section~\ref{sec:empirical:hc:breaks}, we analyze in detail the
French inflation data, extended through 2025.

Table~\ref{tab:emp_hurvich_chen} compares the
original \cite{hurvich-chen-2000} results (for the HC estimator)
with our independent implementations of all five estimators.
For all estimators, we use the same number of frequencies $m$ as in the original study,
where bandwidths were selected based on visual inspection of log-log
periodogram plots.
For the HC-tapered LW estimator, the agreement between our results and the
original estimates is close for all datasets, with differences typically
at the second decimal place or smaller.

In terms of the economic implications of these estimates, there is broad
agreement across estimators.
Global temperatures, with $\hat{d} \in [0.44, 0.50]$, appear to be stationary
with moderate long memory.
The S\&P 500 index appears to have near-unit-root behavior, with $\hat{d} \in [0.97, 0.99]$.
The inflation series reveal heterogeneous persistence across countries:
U.S. inflation has moderate persistence, with $\hat{d} \in [0.56, 0.66]$,
while U.K. inflation has somewhat weaker dependence, with $\hat{d} \in [0.34, 0.49]$.
The five estimators yield widely varying estimates of inflation persistence in France,
with $\hat{d} \in [0.45, 0.82]$.
Because this range includes the nonstationarity boundary at $0.5$, the
estimators disagree on whether the series is even stationary, a case
we examine in detail in Section~\ref{sec:empirical:hc:breaks}.

U.S. real manufacturing wages appear to be nonstationary, with $\hat{d} \in [1.06, 1.56]$
(with the notable exception of ELW), suggesting that wage shocks have permanent effects.
The lower estimate from LW ($\hat{d} = 1.06$) is in line with the known inconsistency of
LW for $d > 1$.
ELW's extremely low estimate ($\hat{d} = 0.06$) reflects sensitivity to the
unknown mean relative to 2ELW and other methods based on tapering.
We will return to this point in the following section.
Similarly, industrial production shows evidence of nonstationarity,
with estimates $\hat{d} \in [1.00, 1.36]$ (again excluding ELW, and with LW being
apparently bounded near unity), indicating that productivity shocks have
persistent effects on growth.
Overall, these results add nuance to the traditional $I(0)$ or $I(1)$
classifications of time series, underscoring the empirical relevance
of fractional integration.

\subsection{Anomalous Memory Estimate for U.S. Industrial Production}
\label{sec:empirical:hc:objective}

\begin{figure}[t!]
\centering
\includegraphics[width=0.9\textwidth]{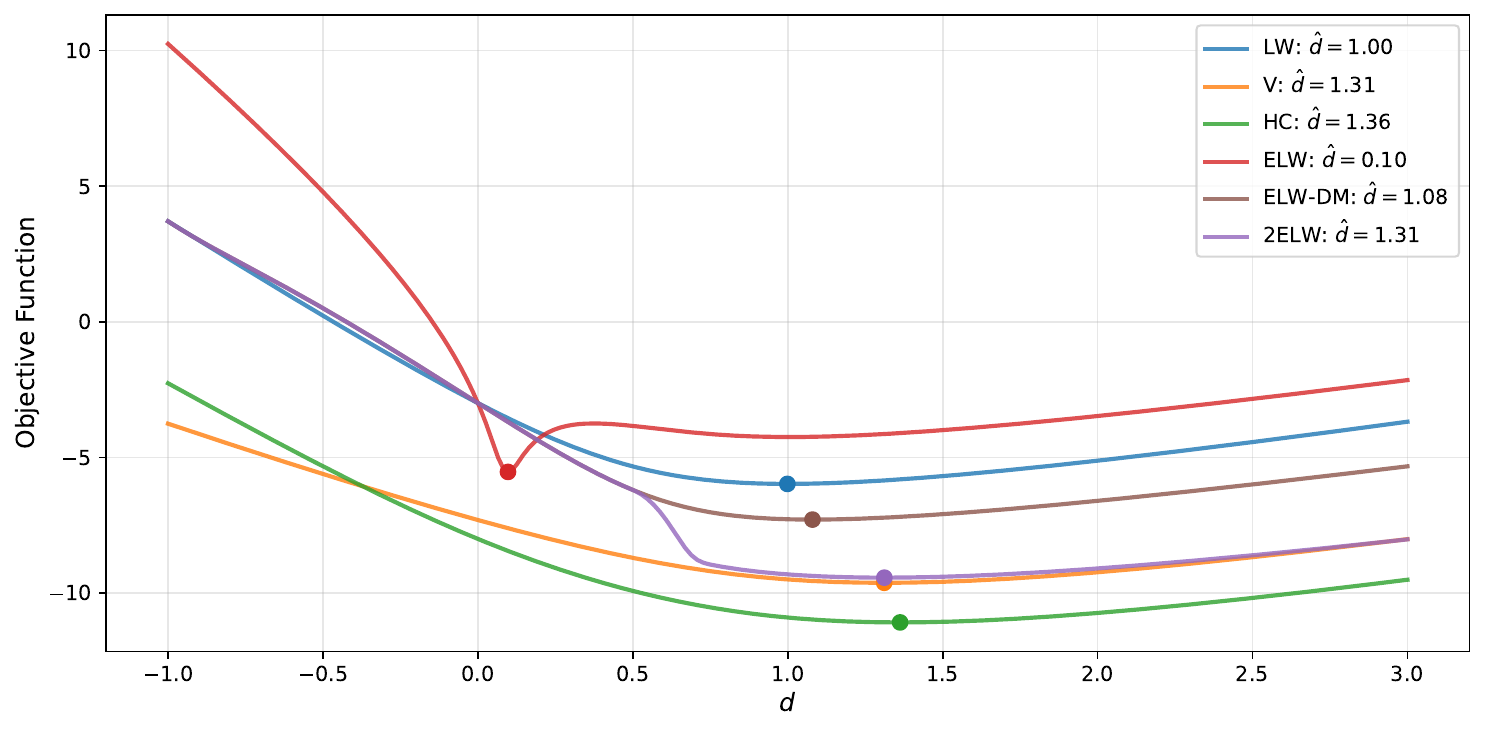}
\caption{Local Whittle objective functions for U.S. industrial
  production (log monthly index, 1957--1997). The LW-based estimators
  (LW, V, HC) have convex objective functions. The ELW objective is
  non-convex with two local minima, its global minimum falling
  at the spurious value $\hat d = 0.10$. The 2ELW and ELW-DM objectives
  coincide for $d \leq \nf{1}{2}$, and 2ELW's adaptive
  mean correction yields a single minimum at $\hat{d} = 1.31$.}
\label{fig:emp_objective}
\end{figure}

We now return to the U.S. industrial production time series, where
ELW yields a spuriously low estimate ($\hat{d} = 0.10$), while the
tapered estimators and 2ELW cluster in $\hat{d} \in [1.31, 1.36]$.
Even the inconsistent untapered LW estimator ($\hat{d} = 1.00$)
yields a more reasonable estimate than ELW.
To understand this difference, we plot the objective functions for
each estimator in Figure~\ref{fig:emp_objective}.
The LW-based objective functions (LW, V, HC) are all convex with a
single global minimum.
The ELW objective, on the other hand, is non-convex and has two local
minima: a spurious global minimum at $\hat{d} = 0.10$ and a shallower
minimum near $d = 1$.
When we apply the ELW estimator to demeaned data, an estimator
discussed in Section 3 of \cite{shimotsu-2010} (denoted ELW-DM
in the figure), the spurious minimum is removed and we obtain
the estimate $\hat{d} = 1.08$.
This suggests that the low estimate produced by ELW is due to the
fact that it assumes the mean is known and equal to zero.

The 2ELW estimator confronts the unknown mean by applying the piecewise
adaptive weight function defined in \eqref{eq:2elw:weight}.
Because $w(d) = 1$ for $d \le 0.5$, the ELW-DM and 2ELW objective
functions coincide exactly up to $d = 0.5$.
After $d = 0.5$, the weight function begins to apply positive weight to
$\hat\mu = X_1$, causing the 2ELW objective function to diverge from
that of ELW-DM.
Because $w(d)$ is continuously differentiable, the transition is smooth
rather than kinked.
The weight shifts fully from $\hat\mu = \bar{X}$ to $\hat\mu = X_1$ by
$d = 0.75$, beyond which the function flattens and attains a single global
minimum at $\hat{d} = 1.31$.

Although a gap between the ELW estimate and the mean-robust
estimates flags possible mean contamination,
the absence of such a gap (agreement across estimators) does not
imply the absence of contamination.
Short-run dynamics, for example, can bias all estimators together
(Section~\ref{sec:mc:comprehensive}).

\subsection{Bandwidth Selection for Daily Values of the S\&P 500 Index}
\label{sec:empirical:hc:bandwidth}

\begin{figure}[t!]
\centering
\includegraphics[width=0.9\textwidth]{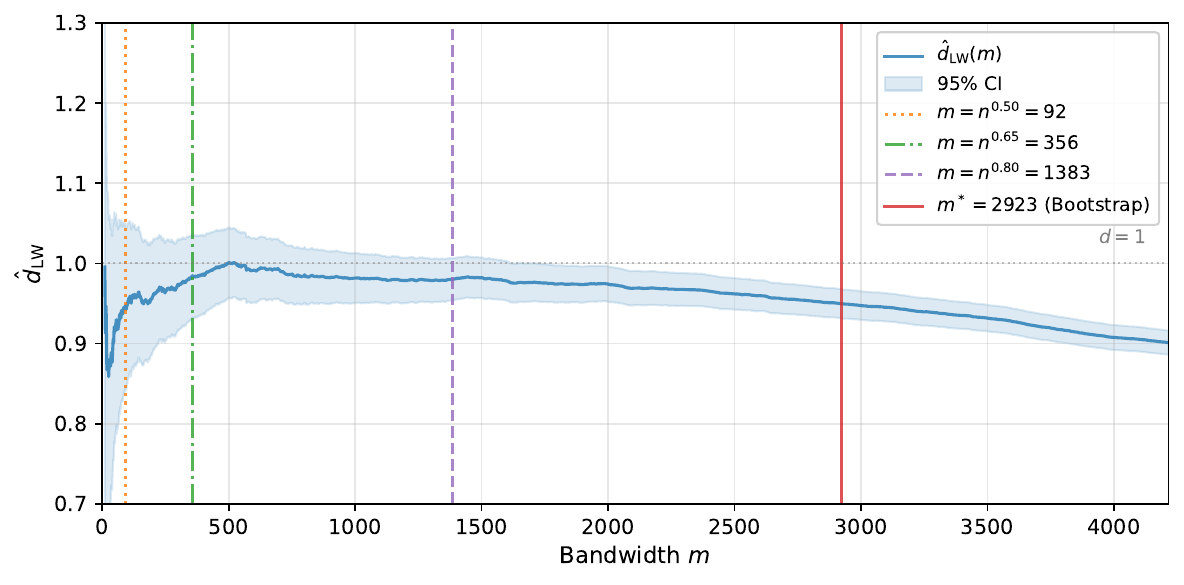}
\caption{Local Whittle estimates by bandwidth using S\&P 500 data. The
  plotted line shows estimates $\hat{d}_{\mathrm{LW}}(m)$ with pointwise
  variance-based bands $\pm 1.96/(2\sqrt{m})$.
  Vertical lines indicate bandwidth choices: the power
  rules $m = n^{0.5}$ (dotted), $m = n^{0.65}$ (dash-dotted), and
  $m = n^{0.8}$ (dashed), and the \cite{arteche-orbe-2017}
  optimal $m^*$ (solid).}
\label{fig:bandwidth_selection}
\end{figure}

We now consider the effects of bandwidth choice $m$.
In applied work it is common to choose
$m = \lfloor n^\alpha \rfloor$ with $\alpha$ ranging from
roughly $0.5$ to $0.8$.
Monte Carlo and empirical work in the literature often uses the
rule $m = \lfloor n^{0.65} \rfloor$ as a baseline.
This rate was used by \cite{shimotsu-phillips-2005} and
\cite{shimotsu-2010} in their simulations and implemented as the
default in both Stata's \code{whittle} command
\citep{baum-hurn-lindsay-2020} and PyELW \citep{pyelw}.
In empirical examples, \cite{shimotsu-2010} used the slightly less
conservative $m = \lfloor n^{0.7} \rfloor$.

\cite{baillie-kapetanios-papailias-2014} proposed cross-validation
approaches focused on forecasting performance, while
\cite{arteche-orbe-2017} developed a bandwidth selection
method based on minimizing a bootstrap approximation of the mean
squared error.
Specifically, an initial LW estimate $\hat{d}$ is used to form the
standardized periodogram $I_X(\lambda_j)\lambda_j^{2\hat{d}}$,
removing the power-law component implied by the memory parameter.
This standardized periodogram is locally resampled to generate bootstrap
samples, and for each candidate bandwidth $m$ the LW estimator is
recomputed to approximate its MSE.
The optimal bandwidth $m^*$ minimizes this bootstrap MSE.

We now illustrate the effects of bandwidth selection using the daily
S\&P 500 index series ($n = 8{,}432$) as an example.
Figure~\ref{fig:bandwidth_selection} plots the LW estimates
$\hat{d}_{\mathrm{LW}}(m)$ as a function of bandwidth $m$, from $m = 10$ to $m = \nf{n}{2}$.
The shaded region shows pointwise 95\% confidence bands
$\hat{d}_{\mathrm{LW}}(m) \pm 1.96/(2\sqrt{m})$.
Estimates range from approximately $0.86$ at extreme values of $m$ to a peak of
approximately $1.0$ near $m=500$.
The confidence intervals narrow as $m$ increases, reflecting the variance
reduction from using additional frequencies.

\begin{table}[t!]
\centering
\begin{threeparttable}
\caption{S\&P 500: Bandwidth Selection Comparison}
\label{tab:emp_bandwidth_selection}
\begin{tabular}{llrrrr}
\toprule
Method & Rule & $m$ & $\hat{d}_{\mathrm{LW}}$ & SE & 95\% CI \\
\midrule
Power Rule & $m = n^{0.50}$ & $92$ & $0.948$ & $0.052$ & $(0.846,1.050)$ \\
 & $m = n^{0.65}$ & $356$ & $0.984$ & $0.026$ & $(0.932,1.036)$ \\
 & $m = n^{0.80}$ & $1{,}383$ & $0.980$ & $0.013$ & $(0.954,1.006)$ \\
\midrule
\cite{arteche-orbe-2017} & Min. Bootstrap MSE & $2{,}923$ & $0.950$ & $0.009$ & $(0.931,0.968)$ \\
\bottomrule
\end{tabular}
\begin{tablenotes}
\footnotesize
\item Notes: Logarithm of S\&P 500 daily stock index value, $n = 8{,}432$. \cite{hurvich-chen-2000}
used $m=n^{0.80}$. Asymptotic standard errors reported with 95\% confidence intervals.
Bootstrap MSE bandwidth selected by minimizing bootstrap MSE with $B = 200$ replications and resampling
width $k_n = 255$ following \cite{arteche-orbe-2017}.
\end{tablenotes}
\end{threeparttable}
\end{table}

The vertical lines in Figure~\ref{fig:bandwidth_selection} indicate different
bandwidth choices. Power rules ($m = n^{0.5}$ through $m = n^{0.8}$) result in
bandwidth choices from $m = 92$ to $m = 1{,}383$,\footnote{To match
  \cite{hurvich-chen-2000}, we use $m = \operatorname{round}(n^\alpha)$ here rather
  than $m = \lfloor n^\alpha \rfloor$, which is used elsewhere in this paper.} which
yields estimates $\hat{d}_{\mathrm{LW}} \in [0.948,0.984]$.
Table~\ref{tab:emp_bandwidth_selection} compares the estimates for different
bandwidth selection methods.
The 95\% confidence intervals from the power rule methods each contain
$d=1.0$, consistent with a unit-root process.

\begin{figure}[t!]
\centering
\includegraphics[width=0.9\textwidth]{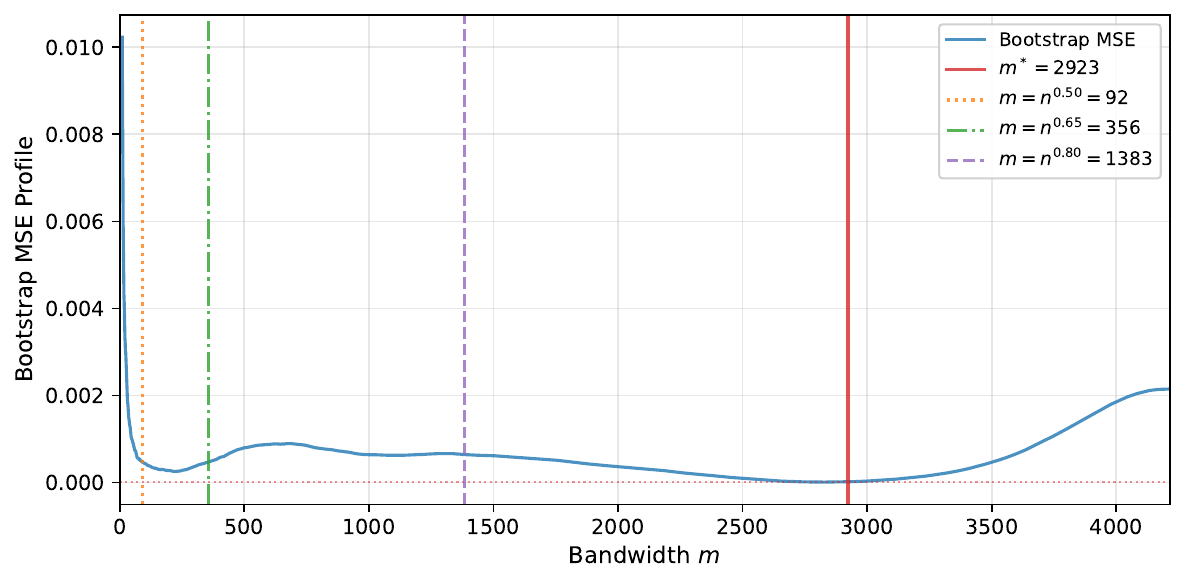}
\caption{Bootstrap MSE profile for S\&P 500 data. The blue line
  represents the estimated MSE for each bandwidth, with the minimum at
  $m^* = 2{,}923$ indicated by the red vertical line. Power rule
  bandwidths are indicated by vertical lines with the same styles as
  in Figure~\ref{fig:bandwidth_selection}.}
\label{fig:bandwidth_mse}
\end{figure}

We also apply the bootstrap MSE minimization procedure of \cite{arteche-orbe-2017},
which selects $m^* = 2{,}923$ and yields $\hat{d}_{\mathrm{LW}} = 0.950$.
We note that this bandwidth is much larger than that of any of the power
rules, which imply conservative choices for this relatively large sample.
This occurs due to the relatively weak short-run dynamics: the resulting
locally standardized periodogram is relatively flat, leading to a large
bandwidth choice $m^*$.
The bootstrap MSE profile, displayed in Figure~\ref{fig:bandwidth_mse},
shows the bias-variance trade-off.
MSE is high at small $m$ due to high variance, then decreases
to a minimum at $m^* = 2{,}923$ before rising again as $m$
approaches $\nf{n}{2}$, where the bias begins to dominate.
Across bandwidths, the estimates are consistent with $d$ at or just below
unity.

\subsection{Structural Breaks and Long Memory in French Inflation}
\label{sec:empirical:hc:breaks}

The wide range of French inflation estimates in
Table~\ref{tab:emp_hurvich_chen}, where $\hat{d} \in [0.45, 0.82]$,
raises the question: which estimator should practitioners trust?
Disagreement of this magnitude, with estimates on both sides of the
nonstationarity boundary $d = 0.5$, has implications for inference,
forecasting, and policy.  One concern is that structural breaks may
inflate long-memory estimates:%
\footnote{Indeed, \cite{caporale-gil-alana-poza-2022} documented
multiple structural breaks in G7 inflation series over 1973--2020.}
\cite{diebold-inoue-2001} and \cite{granger-hyung-2004} showed that
regime-switching and occasional mean shifts mimic the low-frequency
behavior of fractional integration, yielding spuriously positive
estimates of $d$.

A natural approach is to split the sample at break points:
within a stable regime the mean is constant, so within-regime
estimates are free of shift-induced upward bias.%
\footnote{Within-regime estimates remain sensitive to the regime
mean itself, which is a concern for ELW in particular
(Section~\ref{sec:mc:unknown_mean}).}
This yields a useful check: genuine
long memory should remain present in every regime, whereas
spurious memory that is an artifact of level shifts should appear
weaker within regimes \citep{shimotsu-2006-breaks}.

\begin{figure}[t!]
\centering
\includegraphics[width=0.9\textwidth]{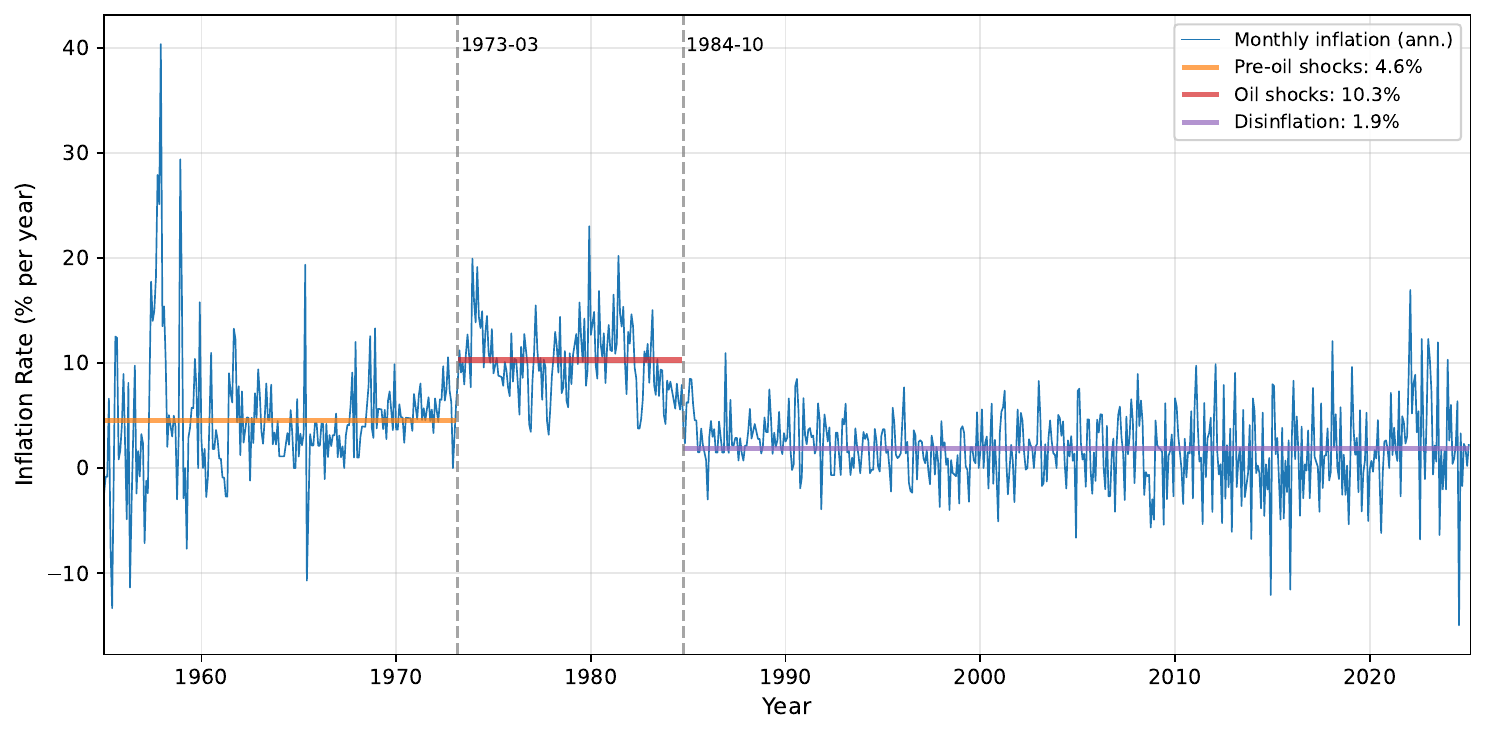}
\caption{French inflation (annualized differences in log monthly CPI)
  with structural breaks. Horizontal lines indicate regime means:
  4.6\%/yr (pre-1973), 10.3\%/yr (1973--1984), and 1.9\%/yr (post-1984);
  within-regime standard deviations are 6.0\%, 3.5\%, and 3.6\%.
  Break dates detected using \cite{bai-perron-2003} methodology.}
\label{fig:structural_breaks}
\end{figure}

Because a structural break analysis is more informative over a long
period, we use an extended French CPI series from January 1955
to March 2025 (longer than the 1957--1997 window used by
\cite{hurvich-chen-2000}), yielding $n = 842$ monthly inflation
observations.\footnote{The series is the consumer price index for
  France (all items) from the OECD Main Economic Indicators, retrieved
  from FRED (series \texttt{FRACPIALLMINMEI}).  Over 1957--1997 it
  coincides exactly with the series used by \cite{hurvich-chen-2000},
  which was drawn from the International Monetary Fund's International
  Financial Statistics database.}
We apply the structural break detection procedure of
\cite{bai-perron-1998,bai-perron-2003}.\footnote{Conventional
  break tests are not robust to long memory and may over-detect
  breaks in its presence.
  \cite{iacone-leybourne-taylor-2013,iacone-leybourne-taylor-2014}
  develop break tests designed to remain valid under
  fractional integration.}
Minimizing the BIC over the full sample selects five breaks, but
that partition results in three segments of only two to three
years, too short for reliable semiparametric estimation.
We therefore focus on the two most prominent breaks, which also
correspond with important economic events: March 1973
(before the first oil shock) and October 1984
(coinciding with the mid-1980s disinflation).
Figure~\ref{fig:structural_breaks} displays the annualized inflation
rates along with the detected break dates and regime-specific means
and standard deviations.

In Table~\ref{tab:french_subsample}, we report the memory parameter
estimates for French inflation using six local Whittle estimators.
The top panel reports estimates for the extended 1955--2025 sample,
including the LWLFC estimator of \cite{hou-perron-2014}.
For comparability with the original analysis and with our replication
in Table~\ref{tab:emp_hurvich_chen}, we retain the bandwidth $m = 40$
chosen by \cite{hurvich-chen-2000}.  On the extended sample this
corresponds to the power rule $m = \lfloor n^{0.55} \rfloor$.
Across the six estimators, full-sample estimates now span
$\hat{d} \in [0.121, 0.801]$.
LWLFC yields the lowest estimate among all estimators, with
$\hat{d} = 0.121$, and the corresponding 95\%
confidence interval contains zero.
This suggests that level shifts are contributing to the
low-frequency power, which the other estimators attribute
to memory.

To investigate this further, in Panels A and B of
Table~\ref{tab:french_subsample}, we report the memory parameter
estimates for each of these regimes with two bandwidth selection
approaches.
Panel~A applies the same implied power rule from the full sample
($m = \lfloor n^{0.55} \rfloor$),
yielding small subsample bandwidths of $m \in \{19, 15, 30\}$.
Panel~B uses the bootstrap-MSE-optimal selection procedure of
\cite{arteche-orbe-2017}, which results in larger subsample
bandwidths of $m \in \{97, 30, 31\}$.
We note that in all panels of this table, for 2ELW we specify
a trend of order zero (i.e., demeaning only).

{\setlength{\tabcolsep}{3.5pt}\begin{table}[t!]
\centering
\begin{threeparttable}
\caption{Inflation in France: Subsample Analysis with Detected Structural Breaks}
\label{tab:french_subsample}
\begin{tabular}{lrrcccccc}
\toprule
Period & $n$ & $m$ & LW & V & HC & ELW & 2ELW & LWLFC \\
\midrule
Full sample & 842 & 40 & $0.495$ & $0.801$ & $0.711$ & $0.518$ & $0.492$ & $0.121$ \\
 & &  & $(0.079)$ & $(0.137)$ & $(0.097)$ & $(0.079)$ & $(0.079)$ & $(0.079)$ \\
\midrule
\multicolumn{9}{c}{\textit{Panel A: Power rule bandwidth ($m = n^{0.55}$)}} \\
\midrule
Pre-oil shocks & 218 & 19 & $0.070$ & $0.087$ & $0.139$ & $0.127$ & $0.090$ & $0.070$ \\
 & &  & $(0.115)$ & $(0.199)$ & $(0.140)$ & $(0.115)$ & $(0.115)$ & $(0.115)$ \\
Oil shocks & 139 & 15 & $0.402$ & $-0.104$ & $0.376$ & $-0.009$ & $0.423$ & $0.402$ \\
 & &  & $(0.129)$ & $(0.224)$ & $(0.158)$ & $(0.129)$ & $(0.129)$ & $(0.129)$ \\
Disinflation & 485 & 30 & $0.350$ & $0.020$ & $0.255$ & $0.360$ & $0.368$ & $0.295$ \\
 & &  & $(0.091)$ & $(0.158)$ & $(0.112)$ & $(0.091)$ & $(0.091)$ & $(0.091)$ \\
\midrule
\multicolumn{9}{c}{\textit{Panel B: Bootstrap-MSE-optimal bandwidth $m^*$}} \\
\midrule
Pre-oil shocks & 218 & 97 & $0.340$ & $0.155$ & $0.341$ & $0.477$ & $0.479$ & $0.340$ \\
 & &  & $(0.051)$ & $(0.088)$ & $(0.062)$ & $(0.051)$ & $(0.051)$ & $(0.051)$ \\
Oil shocks & 139 & 30 & $0.409$ & $0.071$ & $0.387$ & $0.572$ & $0.445$ & $0.409$ \\
 & &  & $(0.091)$ & $(0.158)$ & $(0.112)$ & $(0.091)$ & $(0.091)$ & $(0.091)$ \\
Disinflation & 485 & 31 & $0.348$ & $0.076$ & $0.278$ & $0.360$ & $0.366$ & $0.293$ \\
 & &  & $(0.090)$ & $(0.156)$ & $(0.110)$ & $(0.090)$ & $(0.090)$ & $(0.090)$ \\
\bottomrule
\end{tabular}
\begin{tablenotes}
\footnotesize
\item Notes: Estimates of $d$ with asymptotic standard errors in parentheses. French
inflation is monthly log-differenced CPI from January 1955 to March 2025.
Structural breaks detected following \cite{bai-perron-2003}: March 1973, October 1984. Mean inflation: 4.6\%/yr (pre-1973), 10.3\%/yr (1973--1984), 1.9\%/yr (post-1984).
Panel A uses the same power rule as the full sample.
Panel B uses bootstrap MSE-optimal bandwidth selection \citep{arteche-orbe-2017}.
\end{tablenotes}
\end{threeparttable}
\end{table}
}

The full-sample LW and tapered LW estimates (V and HC)
consistently exceed the subsample estimates.
This is the pattern level shifts induce in estimators that do not
account for a changing mean.
However, the size of the gap differs across estimators.
It is largest for the tapered estimators: in Panel~B, the full-sample HC
estimate ($0.711$) exceeds the subsample estimates by $0.324$--$0.433$,
and the Velasco gaps are larger still.
Tapering provides robustness to polynomial trends and
nonstationarity, but not level shifts.
The Velasco subsample estimates are near zero in both panels, but
the order-three taper discards two-thirds of the periodogram
ordinates, and we therefore treat these estimates as unreliable at
subsample bandwidths rather than as evidence against within-regime
memory.\footnote{Given the Panel~A bandwidths $m \in \{19, 15, 30\}$,
  the order-three Kolmogorov taper retains only $6$, $5$, and
  $10$ periodogram ordinates respectively. The Panel~B bandwidths
  $m \in \{97, 30, 31\}$ leave $32$, $10$, and $10$.}
For the Velasco estimator, the evidence that the level shifts inflate the
estimate comes instead from the full-sample estimate, the largest
among the six and well above the mean-robust 2ELW value of $0.492$.
For the untapered LW estimator the gap is smaller: the full-sample
value ($0.495$) exceeds the subsample estimates ($0.340$, $0.409$,
and $0.348$ in Panel~B) by $0.086$--$0.155$.
The ELW estimate of $-0.009$ for the oil-shocks regime (Panel~A) reflects a
different failure mode: sensitivity to the
regime mean rather than the level shifts themselves.

In Panel~B, the mean-robust 2ELW estimator yields $0.479$, $0.445$,
and $0.366$ across the three regimes, close to its full-sample value
and well above zero.
Excluding Velasco, the other five estimators provide a similar
assessment: they agree to within $0.088$--$0.185$ inside each regime
in Panel~B, and all fifteen of the corresponding 95\% confidence
intervals exclude zero.
This includes the level-shift-robust LWLFC estimator, with
$\hat{d} \in [0.293, 0.409]$ within the regimes, where no level shifts
remain.
Its low full-sample estimate thus reflects the shifts, not
an absence of memory.
Since the estimates remain near $0.4$ across subsamples rather than
collapsing toward zero, this suggests the long memory is genuine
rather than spurious.

The pre-oil-shocks regime illustrates how sensitive subsample
estimates can be to the bandwidth.
We examine this using LW, the estimator for which the bootstrap-MSE
procedure of \cite{arteche-orbe-2017} is designed.
Under the conservative power rule (Panel~A, $m = 19$), the LW estimate
for this regime is $0.070$, with a 95\% confidence interval that
contains zero.  The MSE-optimal bandwidth is much larger ($m = 97$),
and there the estimate is $0.340$, with a confidence interval that
excludes zero.
The apparent absence of memory in this regime is thus specific to the
small bandwidth,\footnote{Starting from the baseline rule
  $m = \lfloor n^{0.65} \rfloor = 33$,
  the LW estimate remains between $0.31$ and $0.38$, with confidence
  intervals excluding zero, at every bandwidth from $33$ to $97$.}
and moderate persistence in line with the other two regimes is
indicated by larger bandwidths.

Formal tests for spurious long memory
\citep{shimotsu-2006-breaks,ohanissian-russell-tsay-2008,qu-2011} test
the null of genuine, constant-memory fractional integration against
shift- or trend-contaminated short memory.
For a series in which the two coexist, a rejection signals the
presence of the shifts, not the absence of genuine memory
\citep[Corollary~1]{qu-2011}.
This is precisely the pattern we find: the \cite{qu-2011} test rejects
the null on the full sample ($W = 1.30$, significant at the 5\% level)
but does not reject in any of the three regimes ($W = 0.65$, $0.64$,
and $0.99$, all below the 10\% critical value of $1.02$), consistent
with level shifts contaminating the full sample and genuine memory
remaining once the regimes are isolated (though non-rejection on
short subsamples is weak evidence).\footnote{Following \cite{qu-2011},
  we use $m = \lceil n^{0.70} \rceil$ and trimming $\varepsilon = 0.05$,
  which is recommended for samples with $n < 500$, as in our subsamples.
  This yields $m = 112$ for the full sample and $m \in \{44, 32, 76\}$
  for the regimes.
  The optional ARFIMA-AIC prewhitening procedure leaves the statistics
  unchanged, which our replication materials confirm using the original
  R code).}

Because long memory and level shifts both concentrate power at
low frequencies, it is difficult to separate them.
We draw two conclusions from the subsample analysis.
First, the regime level shifts appear to be real and to coexist
with long memory.
They nonetheless inflate the mean-sensitive
estimators---the tapered ones most severely---and pull the
LWLFC estimate down to $0.121$.
Second, the evidence is most consistent with genuine long memory of
roughly $0.4$: the mean-robust 2ELW estimate is stable across the
regimes, close to its full-sample value, and does not collapse on
subsamples.
Therefore, the answer to the opening question is not a single
estimator, but rather the comparison exercise that reveals the
level shifts and the memory that survives them.

\section{Guidance for Practitioners}
\label{sec:practical}

This section distills the preceding theoretical, Monte Carlo, and
empirical sections to provide practical recommendations for selecting
and implementing local Whittle estimators in applied work.
Under ideal conditions, every estimator behaves as theory predicts;
the difficulty is that no single one dominates across the scenarios we
considered, so practitioners must match their choice to the features
of their data.

Table~\ref{tab:capability} summarizes the properties that drive this
choice.
It displays each estimator's range of validity, its asymptotic
variance, and its robustness to an unknown mean, a deterministic
trend, and low-frequency level shifts.

\begin{table}[t!]
\centering
\begin{threeparttable}
\caption{Summary of Local Whittle Estimator Properties}
\label{tab:capability}
\begin{tabular}{llcccc}
\toprule
          & Valid range          & Asymptotic          & Unknown   & Time      & Level     \\
Estimator & of $d$               & variance            & mean      & trend     & shifts    \\
\midrule
LW    & $(-\nf{1}{2}, \nf{3}{4})$\tnote{a} & $\nf{1}{4}$        & robust    & sensitive & sensitive \\
V     & $(-\nf{1}{2}, \nf{5}{2})$\tnote{b} & $\nf{3.01}{4}$ & robust    & robust    & sensitive \\
HC    & $(-\nf{1}{2}, \nf{3}{2})$          & $\nf{1.5}{4}$  & robust    & robust    & sensitive \\
ELW   & $(-\infty, \infty)$\tnote{c}     & $\nf{1}{4}$        & sensitive & sensitive & sensitive \\
2ELW  & $(-\nf{1}{2}, 2)$\tnote{d}        & $\nf{1}{4}$        & robust    & robust    & sensitive \\
LWLFC & $[0, \nf{1}{2})$                  & $\nf{1}{4}$        & robust    & sensitive & robust    \\
\bottomrule
\end{tabular}
\begin{tablenotes}\footnotesize
\item[] Notes: ``robust'' denotes consistency without preprocessing while
  ``sensitive'' denotes material bias unless the component is absent or removed.
\item[a] Consistent for $d \in (-\nf{1}{2}, 1)$ and asymptotically normal for
  $d \in (-\nf{1}{2}, \nf{3}{4})$ \citep{velasco-1999}.
\item[b] Kolmogorov taper ($p = 3$), used in our cross-method
  comparisons and empirical applications.
\item[c] Valid for all $d$ provided the optimization interval has width at most
  $\nf{9}{2}$.
\item[d] Range $(-\nf{1}{2}, 2)$ with mean correction, narrowing to
  $(-\nf{1}{2}, \nf{7}{4})$ with detrending.
\end{tablenotes}
\end{threeparttable}
\end{table}

\subsection{Expected Parameter Range}

The first consideration for choosing an estimator should be the
expected parameter range.
If the researcher expects that the process is stationary,
the standard LW and ELW estimators attain the minimal asymptotic
variance of $\nf{1}{4}$.
For applications where low-frequency contamination from level shifts
is suspected, the LWLFC estimator provides robustness without loss of
efficiency, but it is restricted to stationary processes with
$d \in [0, \nf{1}{2})$.
For series where $d$ could lie outside the stationary and invertible
range, practitioners should consider using the ELW and 2ELW estimators.
Tapered LW estimators offer an intermediate solution.
They extend the range of valid parameter values compared to LW,
but at the cost of higher variance.

The choice is more difficult for antipersistent processes.
ELW maintains consistency even for strongly antipersistent processes
in ideal settings, and the Velasco estimator remains accurate in our
simulations well below $d = -\nf{1}{2}$, outside its valid range.
However, the antipersistent region is precisely where ELW is
extremely sensitive to unknown deterministic components,
and the 2ELW estimator fails entirely for $d < -0.6$ in our simulations.
When the researcher suspects a process is antipersistent, we suggest
comparing the ELW and Velasco estimates.
If they agree, it suggests genuine antipersistence.
However, if ELW
produces substantially higher estimates than Velasco, this may
indicate contamination from unknown deterministic components.%
\footnote{The distorting effect of deterministic components on local
  Whittle estimation of the memory parameter is analyzed in detail by
  \cite{iacone-2010}.}

\subsection{Robustness Considerations}

Short-run dynamics affect all local Whittle estimators considered, inducing a
positive bias that grows with the proximity of the autoregressive
root to unity (Section~\ref{sec:mc:comprehensive} and
Appendix~\ref{sec:appendix:robustness}).
Practitioners who suspect
non-trivial short-run dynamics should expect all estimates to be
shifted upward and prefer the exact estimators ELW and 2ELW for their
wider range of validity (though 2ELW should be avoided when strong
antipersistence is suspected).
More conservative bandwidth choices (i.e., smaller values of $m$)
can help mitigate contamination from short-run dynamics
affecting higher frequencies \citep{henry-robinson-1996,henry-2001}.

We saw that the ELW estimator is highly sensitive to
nonzero means and linear trends,
particularly for antipersistent processes
(Sections~\ref{sec:mc:unknown_mean}--\ref{sec:mc:time_trend}).
When analyzing data with potential nonzero means or trends,
practitioners should consider using the 2ELW estimator, noting
that its range is restricted to $d \in (-\nf{1}{2}, \nf{7}{4})$ when detrending.
However, we caution that demeaning or detrending when it is not
necessary can also be detrimental, especially for ELW when $\abs{d}$
is large.

Overall, the tapered estimators and 2ELW are robust to deterministic
components and valid across a wide range of memory parameters, which
makes them the safer default when the properties of the data are
uncertain.
Because LWLFC is valid only within the stationary range, and the
guarantees for LW extend only slightly beyond it, both are best
reserved for series likely to be stationary.
And because ELW is sensitive to deterministic components, it should be
applied to demeaned data or replaced by the 2ELW estimator whenever a
nonzero mean or trend is likely.

\subsection{Bandwidth Selection}

The bandwidth parameter $m$ is an important tuning parameter that
must be selected by the researcher.
Applied work typically adopts power rules $m = \lfloor n^\alpha \rfloor$
with $\alpha$ ranging from roughly $0.5$ to $0.8$, and $\alpha = 0.65$ has emerged as
a practical standard \citep{shimotsu-phillips-2005,shimotsu-2010,baum-hurn-lindsay-2020}.
Data-driven methods such as the bootstrap MSE minimization procedure of
\cite{arteche-orbe-2017} can be used to estimate the optimal bandwidth for LW,
though this procedure often suggests much larger bandwidths than power rules.
We recommend carrying out a sensitivity analysis at multiple bandwidths,
particularly when qualitative conclusions (e.g., $d < 1$ vs.\ $d \geq 1$)
change with reasonable bandwidth choices.

\subsection{Structural Breaks and Spurious Long Memory}

For series prone to regime changes, researchers should
consider structural breaks before interpreting a large estimate of
$d$ as long memory.
Disentangling genuine long memory from spurious long memory faces
difficulties in both directions: semiparametric estimators
overstate $d$ in the presence of breaks
\citep{diebold-inoue-2001,granger-hyung-2004}, while conventional break
tests \citep{bai-perron-1998,bai-perron-2003} are not robust to
long memory and over-detect breaks in its presence, so neither can
be used on its own to test for the other.

Specialized tests have been developed to discriminate between the two.
\cite{shimotsu-2006-breaks} exploits the property that genuine long
memory produces stable $d$ estimates across sample splits.
\cite{ohanissian-russell-tsay-2008} test whether $d$ is invariant
to temporal aggregation, a property satisfied by true long memory
but violated by spurious memory from breaks.
\cite{qu-2011} tests whether the low-frequency spectrum is
consistent with a single memory parameter throughout the band
of frequencies used in estimation, rather than the disproportionate
power at the very lowest frequencies that a level shift or smooth
trend would produce.
\cite{sibbertsen-leschinski-busch-2018} develop a multivariate
version of this test.

Once contamination is suspected, the next step is to determine
the break points, but conventional break-detection procedures
over-detect breaks under long memory (as noted above).
\cite{iacone-leybourne-taylor-2013,iacone-leybourne-taylor-2014}
develop procedures to detect and
date breaks that remain valid under fractional integration.
An informal check is to re-estimate $d$ separately within
each regime.
Although the mean is presumed constant within regimes, it is unknown,
so a mean-robust estimator such as 2ELW may be appropriate.
If the full-sample estimate is positive while the subsample
estimates collapse to zero, this is consistent with spurious
long memory generated by level shifts.
If the subsample estimates instead remain positive and close to
the full-sample value, genuine persistence
may coexist with the shifts (Section~\ref{sec:empirical:hc:breaks}).

As an alternative to testing for breaks, researchers can
turn to estimators designed to be robust to level shifts, such as the trimmed
log-periodogram estimator of \cite{mccloskey-perron-2013} or the LWLFC
estimator of \cite{hou-perron-2014}.
Among local Whittle estimators, LWLFC explicitly
models low-frequency contamination from level shifts, though it
is valid only for stationary processes with $d \in [0, \nf{1}{2})$.

The distinction between genuine and spurious memory carries directly
into forecasting.
When the long memory is genuine, modeling it explicitly through an
ARFIMA specification can improve forecasts over covariance-stationary
short-memory models \citep{granger-joyeux-1980,bhardwaj-swanson-2006}.
Otherwise it should be avoided, since it imposes persistence that
the series will not exhibit out of sample.

\subsection{Multivariate Extensions}

The memory estimates produced by the methods surveyed here are
often used as inputs to a subsequent empirical analysis.
In multivariate settings a natural next question is whether several
fractionally integrated series exhibit common long-run behavior, a
phenomenon known as fractional cointegration.
Semiparametric procedures for detecting it and determining the
cointegrating rank build directly on local Whittle estimation:
\cite{robinson-yajima-2002} develop rank determination for stationary
fractional systems, and
\cite{nielsen-shimotsu-2007} extend the approach to the nonstationary
range using the ELW estimator.
Additionally, the univariate estimates considered here offer useful
guidance on specification and starting values in parametric models,
such as the fractionally cointegrated vector autoregressive model
of \cite{johansen-2008} and \cite{johansen-nielsen-2012}.
Extensions of local Whittle estimation to spatial, functional, and
high-dimensional settings
\citep{chen-wang-2017,li-robinson-shang-2021,baek-duker-pipiras-2023}
likewise take univariate memory estimation as their point of
departure.
The contaminations emphasized above---unknown means, time trends, and
structural breaks---carry over to the multivariate setting, where they
can distort inference about the rank of fractional cointegration.

\subsection{Summary of Recommendations for Empirical Work}

First, we recommend that applied researchers visually inspect their
time series data for obvious trends, level shifts, or regime changes
that may influence the choice of estimator.
Second, in light of the strengths and weaknesses of various local
Whittle estimators, we recommend applying multiple methods to
estimate $d$ as a robustness check: agreement across methods with
different robustness properties lends confidence, whereas disagreement
is a caution that one or more methods' assumptions may not hold.
Agreement does not guarantee correctness, however, since a common
contamination such as persistent short-run dynamics can bias all
methods together.
Third, we recommend conducting a bandwidth sensitivity analysis across
$\alpha \in \{0.5, 0.6, 0.65, 0.7, 0.8\}$.
Researchers should also consider using the
bootstrap MSE-optimal selection procedure of \cite{arteche-orbe-2017}.
Fourth, when estimators disagree or visual inspection suggests regime changes,
apply formal break detection methods and robust estimation of $d$
as described above.

\section{Conclusion}
\label{sec:conclusion}

This paper carries out a systematic evaluation of six local Whittle
methods for estimating the memory parameter $d$,
reproducing key Monte Carlo results from the literature and extending
them with cross-method comparisons under short-run dynamics, unknown
means, deterministic time trends, and non-Gaussian innovations.
The simulations and the accompanying empirical case studies document
each estimator's characteristic failure modes.

As the preceding analysis shows, no single local Whittle
estimator dominates in all settings.
The ELW estimator attains the minimal asymptotic variance of
$\nf{1}{4}$ under ideal conditions, but is sensitive to deterministic
components.
Tapered methods offer robustness to such components
at the cost of increased variance and a more limited
range of valid parameter values than ELW.
The 2ELW estimator includes adaptive mean correction
and detrending, also at the expense of the parameter
range.
Data-driven bandwidth selection for the exact estimators remains an
open problem.

Our recommendations are based on combining existing theory with
simulation evidence, and the diagnostics we suggest are informal.
Because no single method is universally robust,
we recommend computing several estimators with complementary
strengths, matching the choice to the features one suspects in the
data, and interpreting any disagreement among them in terms of
possible contaminations.

\bibliographystyle{chicago}
\bibliography{lws}

\newpage
\appendix

\section{Robustness of the Estimator Comparison}
\label{sec:appendix:robustness}

The recommendations in Section~\ref{sec:practical} are based on
the cross-method comparison of Section~\ref{sec:mc}, where the
sample size, the form of the short-run dynamics, and the innovation
distribution were all fixed.
In this appendix, we consider robustness to these choices. We hold
the overall simulation design of Section~\ref{sec:mc:comprehensive}
fixed and vary, in turn, the sample size, the form of the
short-run dynamics, and the innovation distribution.
As before, each estimator is evaluated across
the full $d$ grid, including values outside the respective regions
of validity.

\subsection{Sensitivity to Sample Size}
\label{sec:appendix:n}

Tables~\ref{tab:mc_comp_n250} and~\ref{tab:mc_comp_n1000} repeat the
comprehensive comparison at $n = 250$ and $n = 1000$, halving and
doubling the baseline sample size $n = 500$ while using the same
$m = \lfloor n^{0.65} \rfloor$ bandwidth rule.
The qualitative ranking of Section~\ref{sec:mc:comprehensive} is
preserved at both sample sizes: ELW (and 2ELW within its range)
remains valid and uniformly accurate across the full $d$ grid, while
the standard LW and LWLFC estimators continue to fail once $d$
leaves their region of validity.
Within each estimator's valid range, bias and MSE decline with $n$ as
expected.
Overall, the patterns documented in the main text are stable across
the sample sizes considered.

\subsection{Sensitivity to the Short-Run Specification}
\label{sec:appendix:shortrun}

In Table~\ref{tab:mc_comp_ma1}, we replace the AR(1) short-run
component with an MA(1) component, simulating $\ARFIMA(0,d,1)$
processes with moving-average parameter $\theta \in \{0.5, 0.8\}$
at $n = 500$. (The baseline case $\theta = 0$ coincides with the
$\rho = 0$ panel of Table~\ref{tab:mc_comprehensive} and is
omitted.)
Even though the MA(1) term results in a different spectral density
shape near the origin, the overall contamination patterns carry
over.
The ordering of the estimators under MA(1) dynamics matches the
AR(1) ordering, with the exact methods most accurate overall, the
Velasco estimator carrying the largest bias among the valid methods,
and LW and LWLFC failing outside their valid range.
The effects are, however, milder by more than an order of
magnitude: at $d = 0$ with $\theta = 0.8$ the ELW bias is only
$0.015$, compared to $0.416$ for the AR(1) case at $\rho = 0.8$.

Finally, in Table~\ref{tab:mc_comp_arma11} we consider
ARMA(1,1) short-run dynamics over three specifications of
increasing difficulty:
$(\phi,\theta) \in \{(0.5,0.5),(0.8,0.5),(0.9,0.5)\}$,
with the last being a near-unit-root case.
This is a much more challenging design than in the main text:
the ELW bias at $d = 0$ rises from $0.116$ to $0.434$ to
$0.683$ across the three configurations.
The contrast with the negligible MA(1) effects indicates that
the severity of short-run bias is governed by the proximity
of the autoregressive root to unity, not by the particular
short-run model.
The range and failure structure are essentially preserved: at
$d = 2.2$ under $(\phi,\theta) = (0.9,0.5)$ the LW and LWLFC
estimators collapse toward the unit-root boundary (bias near
$-1.15$), while the exact methods continue to track $d$ in the
nonstationary region, displaced by the same uniform
bias seen at central $d$.
Short-run dynamics bias the estimators in unison, whereas
mean or trend contamination drives ELW away from the
tapered estimators.

\subsection{Sensitivity to the Innovation Distribution}
\label{sec:appendix:innovations}

The experiments above all use Gaussian innovations, but many series to
which these estimators are applied, such as asset returns, exhibit
heavy tails and volatility clustering.
Therefore, as a robustness check, we repeat the baseline
$\ARFIMA(0,d,0)$ comparison with both $t(5)$ and GARCH(1,1)
innovations, each standardized to have mean zero and unit
variance.
The GARCH(1,1) innovations satisfy
$\varepsilon_t = \sigma_t \eta_t$ with
$\sigma_t^2 = \omega + \alpha \varepsilon_{t-1}^2 + \beta \sigma_{t-1}^2$,
$\eta_t \sim \text{i.i.d.}\ \Normal(0,1)$, and
$(\omega, \alpha, \beta) = (0.1, 0.1, 0.8)$, so that volatility is
persistent ($\alpha + \beta = 0.9$).
For the LW estimator, \cite{robinson-henry-1999} showed that
conditional heteroskedasticity of this form leaves the
$\Normal(0, \nf{1}{4})$ limiting distribution unchanged.

The results, displayed in Table~\ref{tab:mc_comp_heavy}, indicate that
the comparison of Section~\ref{sec:mc:comprehensive} does not depend
on Gaussianity specifically.
Relative to the Gaussian baseline (the $\rho = 0$ panel of
Table~\ref{tab:mc_comprehensive}), heavy-tailed $t(5)$
innovations (top panel) leave both bias and MSE essentially
unchanged.
The conditional heteroskedasticity of GARCH(1,1) innovations (bottom
panel) likewise produces similar biases and only a slight increase in
variance, consistent with the findings of \cite{robinson-henry-1999}.
The characteristic failure modes of the baseline are preserved in both
panels:
LW and LWLFC collapse toward the unit-root boundary for $d \geq 1.8$,
and 2ELW fails under strong antipersistence.

{\setlength{\tabcolsep}{3.5pt}
\begin{table}[tp]
\centering
\begin{threeparttable}
\caption{Estimator Comparison at $n=250$}
\label{tab:mc_comp_n250}
\scriptsize
\begin{tabular}{cc|rrrrrr|rrrrrr}
\toprule
&  & \multicolumn{6}{c|}{Bias} & \multicolumn{6}{c}{MSE} \\
\cmidrule(lr){3-8} \cmidrule(lr){9-14}
$d$ & $\rho$ & LW & V & HC & ELW & 2ELW & LWLFC & LW & V & HC & ELW & 2ELW & LWLFC \\
\midrule
-2.2 & 0.0 &   1.387 &   0.050 &   0.828 &  -0.000 &   1.109 &   1.231 & \largemse{  2.035} & \largemse{  0.051} & \largemse{  0.770} &   0.011 & \largemse{  1.306} & \largemse{  1.525} \\
-1.8 & 0.0 &   0.907 &   0.025 &   0.446 &   0.000 &   0.724 &   0.817 & \largemse{  0.917} &   0.049 & \largemse{  0.248} &   0.011 & \largemse{  0.571} & \largemse{  0.672} \\
-1.2 & 0.0 &   0.283 &   0.002 &   0.112 &  -0.002 &   0.278 &   0.233 & \largemse{  0.119} &   0.048 &   0.030 &   0.011 & \largemse{  0.099} & \largemse{  0.060} \\
-0.6 & 0.0 &   0.021 &  -0.016 &   0.036 &  -0.001 &   0.004 &  -0.018 &   0.012 &   0.048 &   0.018 &   0.011 &   0.010 &   0.015 \\
-0.3 & 0.0 &  -0.003 &  -0.025 &   0.018 &  -0.001 &  -0.004 &  -0.038 &   0.011 &   0.049 &   0.017 &   0.011 &   0.011 &   0.017 \\
 0.0 & 0.0 &  -0.011 &  -0.029 &   0.006 &  -0.002 &  -0.002 &  -0.052 &   0.011 &   0.049 &   0.017 &   0.011 &   0.011 &   0.020 \\
 0.3 & 0.0 &  -0.011 &  -0.028 &  -0.002 &  -0.002 &   0.001 &  -0.075 &   0.011 & \largemse{  0.050} &   0.017 &   0.011 &   0.011 &   0.032 \\
 0.6 & 0.0 &  -0.002 &  -0.022 &  -0.007 &  -0.001 &   0.016 &  -0.131 &   0.011 &   0.049 &   0.017 &   0.011 &   0.010 & \largemse{  0.101} \\
 1.2 & 0.0 &  -0.128 &  -0.011 &  -0.012 &  -0.001 &  -0.001 &  -0.132 &   0.025 &   0.049 &   0.017 &   0.010 &   0.010 &   0.043 \\
 1.8 & 0.0 &  -0.723 &   0.022 &   0.004 &  -0.002 &  -0.002 &  -0.712 & \largemse{  0.545} &   0.050 &   0.016 &   0.011 &   0.011 & \largemse{  0.543} \\
 2.2 & 0.0 &  -1.151 &   0.056 &   0.047 &  -0.003 &  -0.003 &  -1.141 & \largemse{  1.342} & \largemse{  0.053} &   0.018 &   0.010 &   0.010 & \largemse{  1.334} \\
\midrule
-2.2 & 0.5 &   1.223 &   0.234 &   0.769 &   0.154 &   1.047 &   1.206 & \largemse{  1.596} & \largemse{  0.103} & \largemse{  0.655} &   0.035 & \largemse{  1.155} & \largemse{  1.455} \\
-1.8 & 0.5 &   0.776 &   0.219 &   0.460 &   0.156 &   0.715 &   0.806 & \largemse{  0.676} & \largemse{  0.096} & \largemse{  0.241} &   0.035 & \largemse{  0.546} & \largemse{  0.651} \\
-1.2 & 0.5 &   0.299 &   0.192 &   0.254 &   0.157 &   0.303 &   0.272 & \largemse{  0.110} & \largemse{  0.086} & \largemse{  0.081} &   0.035 & \largemse{  0.106} & \largemse{  0.082} \\
-0.6 & 0.5 &   0.162 &   0.175 &   0.204 &   0.155 &   0.154 &   0.148 &   0.038 & \largemse{  0.081} & \largemse{  0.060} &   0.036 &   0.035 &   0.035 \\
-0.3 & 0.5 &   0.150 &   0.171 &   0.192 &   0.157 &   0.155 &   0.137 &   0.033 & \largemse{  0.077} & \largemse{  0.054} &   0.035 &   0.035 &   0.032 \\
 0.0 & 0.5 &   0.144 &   0.168 &   0.180 &   0.156 &   0.157 &   0.130 &   0.031 & \largemse{  0.077} &   0.050 &   0.035 &   0.035 &   0.030 \\
 0.3 & 0.5 &   0.141 &   0.167 &   0.170 &   0.154 &   0.161 &   0.119 &   0.030 & \largemse{  0.077} &   0.047 &   0.035 &   0.038 &   0.031 \\
 0.6 & 0.5 &   0.148 &   0.171 &   0.167 &   0.156 &   0.159 &   0.099 &   0.033 & \largemse{  0.079} &   0.045 &   0.035 &   0.035 & \largemse{  0.060} \\
 1.2 & 0.5 &  -0.081 &   0.179 &   0.161 &   0.155 &   0.155 &  -0.080 &   0.023 & \largemse{  0.081} &   0.044 &   0.035 &   0.035 &   0.025 \\
 1.8 & 0.5 &  -0.721 &   0.213 &   0.176 &   0.157 &   0.157 &  -0.713 & \largemse{  0.545} & \largemse{  0.093} &   0.047 &   0.035 &   0.035 & \largemse{  0.543} \\
 2.2 & 0.5 &  -1.151 &   0.249 &   0.216 &   0.157 &   0.156 &  -1.143 & \largemse{  1.344} & \largemse{  0.111} & \largemse{  0.062} &   0.035 &   0.035 & \largemse{  1.337} \\
\midrule
-2.2 & 0.8 &   1.201 &   0.633 &   0.906 &   0.521 &   1.143 &   1.210 & \largemse{  1.518} & \largemse{  0.453} & \largemse{  0.858} & \largemse{  0.285} & \largemse{  1.342} & \largemse{  1.466} \\
-1.8 & 0.8 &   0.826 &   0.617 &   0.710 &   0.523 &   0.843 &   0.831 & \largemse{  0.722} & \largemse{  0.431} & \largemse{  0.522} & \largemse{  0.286} & \largemse{  0.734} & \largemse{  0.696} \\
-1.2 & 0.8 &   0.562 &   0.596 &   0.613 &   0.523 &   0.545 &   0.555 & \largemse{  0.329} & \largemse{  0.405} & \largemse{  0.393} & \largemse{  0.286} & \largemse{  0.307} & \largemse{  0.322} \\
-0.6 & 0.8 &   0.515 &   0.582 &   0.582 &   0.522 &   0.521 &   0.513 & \largemse{  0.278} & \largemse{  0.389} & \largemse{  0.358} & \largemse{  0.286} & \largemse{  0.285} & \largemse{  0.276} \\
-0.3 & 0.8 &   0.508 &   0.581 &   0.572 &   0.522 &   0.523 &   0.506 & \largemse{  0.270} & \largemse{  0.388} & \largemse{  0.347} & \largemse{  0.286} & \largemse{  0.287} & \largemse{  0.269} \\
 0.0 & 0.8 &   0.503 &   0.576 &   0.562 &   0.522 &   0.532 &   0.501 & \largemse{  0.265} & \largemse{  0.382} & \largemse{  0.336} & \largemse{  0.286} & \largemse{  0.297} & \largemse{  0.264} \\
 0.3 & 0.8 &   0.499 &   0.577 &   0.555 &   0.523 &   0.524 &   0.495 & \largemse{  0.262} & \largemse{  0.383} & \largemse{  0.328} & \largemse{  0.286} & \largemse{  0.287} & \largemse{  0.261} \\
 0.6 & 0.8 &   0.474 &   0.581 &   0.550 &   0.523 &   0.523 &   0.472 & \largemse{  0.237} & \largemse{  0.389} & \largemse{  0.323} & \largemse{  0.286} & \largemse{  0.286} & \largemse{  0.238} \\
 1.2 & 0.8 &  -0.021 &   0.592 &   0.545 &   0.522 &   0.522 &  -0.018 &   0.041 & \largemse{  0.401} & \largemse{  0.317} & \largemse{  0.285} & \largemse{  0.285} &   0.044 \\
 1.8 & 0.8 &  -0.720 &   0.621 &   0.557 &   0.522 &   0.521 &  -0.711 & \largemse{  0.548} & \largemse{  0.434} & \largemse{  0.329} & \largemse{  0.286} & \largemse{  0.284} & \largemse{  0.546} \\
 2.2 & 0.8 &  -1.152 &   0.637 &   0.584 &   0.522 &   0.520 &  -1.145 & \largemse{  1.346} & \largemse{  0.455} & \largemse{  0.357} & \largemse{  0.285} & \largemse{  0.283} & \largemse{  1.340} \\
\bottomrule
\end{tabular}
\begin{tablenotes}
\footnotesize
\item Notes: Monte Carlo results for $\ARFIMA(1,d,0)$ processes with $n=250$, $m=36$, 10,000 replications.
    Shaded cells indicate $\text{MSE} > 0.05$.
\item LW = Local Whittle, V = Velasco (Kolmogorov), HC = Hurvich--Chen, ELW = Exact Local Whittle, 2ELW = Two-step ELW, LWLFC = Local Whittle robust to low-frequency contamination.
\end{tablenotes}
\end{threeparttable}
\end{table}

\begin{table}[tp]
\centering
\begin{threeparttable}
\caption{Estimator Comparison at $n=1000$}
\label{tab:mc_comp_n1000}
\scriptsize
\begin{tabular}{cc|rrrrrr|rrrrrr}
\toprule
&  & \multicolumn{6}{c|}{Bias} & \multicolumn{6}{c}{MSE} \\
\cmidrule(lr){3-8} \cmidrule(lr){9-14}
$d$ & $\rho$ & LW & V & HC & ELW & 2ELW & LWLFC & LW & V & HC & ELW & 2ELW & LWLFC \\
\midrule
-2.2 & 0.0 &   1.593 &   0.046 &   1.117 &  -0.001 &   1.266 &   1.295 & \largemse{  2.617} &   0.016 & \largemse{  1.321} &   0.004 & \largemse{  1.668} & \largemse{  1.710} \\
-1.8 & 0.0 &   1.065 &   0.035 &   0.640 &  -0.002 &   0.795 &   0.829 & \largemse{  1.208} &   0.015 & \largemse{  0.466} &   0.004 & \largemse{  0.682} & \largemse{  0.695} \\
-1.2 & 0.0 &   0.345 &   0.024 &   0.120 &  -0.003 &   0.257 &   0.224 & \largemse{  0.156} &   0.014 &   0.025 &   0.004 & \largemse{  0.080} & \largemse{  0.054} \\
-0.6 & 0.0 &   0.017 &   0.019 &   0.016 &  -0.002 &   0.002 &  -0.001 &   0.004 &   0.014 &   0.006 &   0.004 &   0.003 &   0.004 \\
-0.3 & 0.0 &  -0.000 &   0.016 &   0.007 &  -0.001 &  -0.002 &  -0.015 &   0.004 &   0.014 &   0.006 &   0.004 &   0.004 &   0.004 \\
 0.0 & 0.0 &  -0.004 &   0.015 &   0.002 &  -0.002 &  -0.002 &  -0.022 &   0.004 &   0.014 &   0.005 &   0.004 &   0.004 &   0.005 \\
 0.3 & 0.0 &  -0.004 &   0.017 &  -0.002 &  -0.002 &  -0.001 &  -0.029 &   0.004 &   0.014 &   0.006 &   0.004 &   0.004 &   0.007 \\
 0.6 & 0.0 &   0.004 &   0.015 &  -0.006 &  -0.003 &   0.010 &  -0.049 &   0.004 &   0.014 &   0.005 &   0.003 &   0.003 &   0.017 \\
 1.2 & 0.0 &  -0.125 &   0.025 &  -0.006 &  -0.002 &  -0.002 &  -0.125 &   0.021 &   0.014 &   0.005 &   0.004 &   0.004 &   0.021 \\
 1.8 & 0.0 &  -0.745 &   0.048 &   0.005 &  -0.001 &  -0.001 &  -0.744 & \largemse{  0.569} &   0.016 &   0.005 &   0.003 &   0.003 & \largemse{  0.568} \\
 2.2 & 0.0 &  -1.168 &   0.064 &   0.028 &  -0.002 &  -0.002 &  -1.167 & \largemse{  1.373} &   0.019 &   0.006 &   0.004 &   0.004 & \largemse{  1.372} \\
\midrule
-2.2 & 0.5 &   1.401 &   0.122 &   0.997 &   0.065 &   1.150 &   1.208 & \largemse{  2.038} &   0.029 & \largemse{  1.062} &   0.008 & \largemse{  1.383} & \largemse{  1.461} \\
-1.8 & 0.5 &   0.895 &   0.113 &   0.553 &   0.065 &   0.705 &   0.802 & \largemse{  0.866} &   0.027 & \largemse{  0.351} &   0.008 & \largemse{  0.538} & \largemse{  0.643} \\
-1.2 & 0.5 &   0.274 &   0.102 &   0.144 &   0.066 &   0.255 &   0.220 & \largemse{  0.096} &   0.024 &   0.027 &   0.008 & \largemse{  0.076} & \largemse{  0.050} \\
-0.6 & 0.5 &   0.075 &   0.094 &   0.085 &   0.065 &   0.066 &   0.066 &   0.010 &   0.023 &   0.013 &   0.008 &   0.008 &   0.009 \\
-0.3 & 0.5 &   0.065 &   0.093 &   0.078 &   0.065 &   0.065 &   0.058 &   0.008 &   0.023 &   0.012 &   0.008 &   0.008 &   0.007 \\
 0.0 & 0.5 &   0.062 &   0.092 &   0.073 &   0.065 &   0.065 &   0.054 &   0.007 &   0.022 &   0.011 &   0.008 &   0.008 &   0.007 \\
 0.3 & 0.5 &   0.062 &   0.094 &   0.070 &   0.065 &   0.066 &   0.052 &   0.007 &   0.023 &   0.011 &   0.008 &   0.008 &   0.007 \\
 0.6 & 0.5 &   0.069 &   0.093 &   0.066 &   0.065 &   0.070 &   0.050 &   0.009 &   0.023 &   0.010 &   0.008 &   0.008 &   0.012 \\
 1.2 & 0.5 &  -0.105 &   0.103 &   0.068 &   0.065 &   0.065 &  -0.105 &   0.019 &   0.024 &   0.010 &   0.008 &   0.008 &   0.020 \\
 1.8 & 0.5 &  -0.747 &   0.118 &   0.074 &   0.065 &   0.065 &  -0.747 & \largemse{  0.572} &   0.028 &   0.011 &   0.008 &   0.008 & \largemse{  0.572} \\
 2.2 & 0.5 &  -1.167 &   0.137 &   0.097 &   0.066 &   0.066 &  -1.167 & \largemse{  1.372} &   0.033 &   0.015 &   0.008 &   0.008 & \largemse{  1.371} \\
\midrule
-2.2 & 0.8 &   1.307 &   0.407 &   0.988 &   0.326 &   1.128 &   1.203 & \largemse{  1.776} & \largemse{  0.181} & \largemse{  1.030} & \largemse{  0.111} & \largemse{  1.319} & \largemse{  1.447} \\
-1.8 & 0.8 &   0.841 &   0.399 &   0.608 &   0.326 &   0.746 &   0.804 & \largemse{  0.752} & \largemse{  0.175} & \largemse{  0.393} & \largemse{  0.111} & \largemse{  0.579} & \largemse{  0.647} \\
-1.2 & 0.8 &   0.402 &   0.390 &   0.376 &   0.327 &   0.387 &   0.391 & \largemse{  0.169} & \largemse{  0.167} & \largemse{  0.147} & \largemse{  0.111} & \largemse{  0.155} & \largemse{  0.160} \\
-0.6 & 0.8 &   0.327 &   0.383 &   0.352 &   0.325 &   0.325 &   0.327 & \largemse{  0.111} & \largemse{  0.162} & \largemse{  0.130} & \largemse{  0.110} & \largemse{  0.110} & \largemse{  0.111} \\
-0.3 & 0.8 &   0.323 &   0.381 &   0.348 &   0.326 &   0.326 &   0.322 & \largemse{  0.108} & \largemse{  0.161} & \largemse{  0.128} & \largemse{  0.110} & \largemse{  0.110} & \largemse{  0.108} \\
 0.0 & 0.8 &   0.321 &   0.380 &   0.344 &   0.326 &   0.326 &   0.321 & \largemse{  0.107} & \largemse{  0.160} & \largemse{  0.125} & \largemse{  0.110} & \largemse{  0.110} & \largemse{  0.107} \\
 0.3 & 0.8 &   0.321 &   0.382 &   0.342 &   0.326 &   0.330 &   0.321 & \largemse{  0.107} & \largemse{  0.161} & \largemse{  0.123} & \largemse{  0.111} & \largemse{  0.113} & \largemse{  0.107} \\
 0.6 & 0.8 &   0.321 &   0.385 &   0.342 &   0.327 &   0.327 &   0.311 & \largemse{  0.107} & \largemse{  0.163} & \largemse{  0.123} & \largemse{  0.111} & \largemse{  0.111} & \largemse{  0.119} \\
 1.2 & 0.8 &  -0.054 &   0.388 &   0.339 &   0.326 &   0.326 &  -0.053 &   0.026 & \largemse{  0.166} & \largemse{  0.121} & \largemse{  0.110} & \largemse{  0.110} &   0.026 \\
 1.8 & 0.8 &  -0.742 &   0.401 &   0.344 &   0.327 &   0.326 &  -0.741 & \largemse{  0.568} & \largemse{  0.175} & \largemse{  0.124} & \largemse{  0.111} & \largemse{  0.111} & \largemse{  0.568} \\
 2.2 & 0.8 &  -1.165 &   0.415 &   0.358 &   0.326 &   0.326 &  -1.165 & \largemse{  1.368} & \largemse{  0.187} & \largemse{  0.134} & \largemse{  0.111} & \largemse{  0.111} & \largemse{  1.367} \\
\bottomrule
\end{tabular}
\begin{tablenotes}
\footnotesize
\item Notes: Monte Carlo results for $\ARFIMA(1,d,0)$ processes with $n=1000$, $m=89$, 10,000 replications.
    Shaded cells indicate $\text{MSE} > 0.05$.
\item LW = Local Whittle, V = Velasco (Kolmogorov), HC = Hurvich--Chen, ELW = Exact Local Whittle, 2ELW = Two-step ELW, LWLFC = Local Whittle robust to low-frequency contamination.
\end{tablenotes}
\end{threeparttable}
\end{table}

\begin{table}[tp]
\centering
\begin{threeparttable}
\caption{Estimator Comparison under MA(1) Short-Run Dynamics}
\label{tab:mc_comp_ma1}
\scriptsize
\begin{tabular}{cc|rrrrrr|rrrrrr}
\toprule
&  & \multicolumn{6}{c|}{Bias} & \multicolumn{6}{c}{MSE} \\
\cmidrule(lr){3-8} \cmidrule(lr){9-14}
$d$ & $\theta$ & LW & V & HC & ELW & 2ELW & LWLFC & LW & V & HC & ELW & 2ELW & LWLFC \\
\midrule
-2.2 & 0.5 &   1.352 &   0.089 &   0.901 &   0.013 &   1.109 &   1.208 & \largemse{  1.916} &   0.034 & \largemse{  0.888} &   0.006 & \largemse{  1.301} & \largemse{  1.461} \\
-1.8 & 0.5 &   0.872 &   0.076 &   0.483 &   0.014 &   0.694 &   0.803 & \largemse{  0.835} &   0.032 & \largemse{  0.281} &   0.006 & \largemse{  0.527} & \largemse{  0.645} \\
-1.2 & 0.5 &   0.255 &   0.058 &   0.108 &   0.014 &   0.247 &   0.216 & \largemse{  0.093} &   0.029 &   0.022 &   0.006 & \largemse{  0.077} &   0.049 \\
-0.6 & 0.5 &   0.029 &   0.047 &   0.040 &   0.013 &   0.016 &   0.006 &   0.007 &   0.029 &   0.011 &   0.006 &   0.006 &   0.008 \\
-0.3 & 0.5 &   0.013 &   0.045 &   0.029 &   0.013 &   0.012 &  -0.006 &   0.006 &   0.029 &   0.010 &   0.006 &   0.006 &   0.008 \\
 0.0 & 0.5 &   0.008 &   0.042 &   0.020 &   0.013 &   0.013 &  -0.015 &   0.006 &   0.028 &   0.010 &   0.006 &   0.006 &   0.009 \\
 0.3 & 0.5 &   0.008 &   0.042 &   0.014 &   0.013 &   0.014 &  -0.027 &   0.006 &   0.028 &   0.010 &   0.006 &   0.006 &   0.012 \\
 0.6 & 0.5 &   0.018 &   0.044 &   0.010 &   0.014 &   0.028 &  -0.052 &   0.007 &   0.028 &   0.010 &   0.006 &   0.006 &   0.035 \\
 1.2 & 0.5 &  -0.120 &   0.056 &   0.008 &   0.012 &   0.012 &  -0.120 &   0.022 &   0.030 &   0.010 &   0.006 &   0.006 &   0.024 \\
 1.8 & 0.5 &  -0.737 &   0.083 &   0.022 &   0.014 &   0.014 &  -0.733 & \largemse{  0.560} &   0.034 &   0.010 &   0.006 &   0.006 & \largemse{  0.557} \\
 2.2 & 0.5 &  -1.161 &   0.109 &   0.054 &   0.013 &   0.013 &  -1.158 & \largemse{  1.360} &   0.039 &   0.012 &   0.006 &   0.006 & \largemse{  1.356} \\
\midrule
-2.2 & 0.8 &   1.317 &   0.087 &   0.878 &   0.014 &   1.091 &   1.204 & \largemse{  1.819} &   0.034 & \largemse{  0.846} &   0.006 & \largemse{  1.260} & \largemse{  1.451} \\
-1.8 & 0.8 &   0.839 &   0.075 &   0.465 &   0.016 &   0.677 &   0.802 & \largemse{  0.777} &   0.033 & \largemse{  0.262} &   0.006 & \largemse{  0.504} & \largemse{  0.643} \\
-1.2 & 0.8 &   0.245 &   0.060 &   0.106 &   0.017 &   0.242 &   0.215 & \largemse{  0.087} &   0.031 &   0.021 &   0.006 & \largemse{  0.074} &   0.048 \\
-0.6 & 0.8 &   0.029 &   0.046 &   0.040 &   0.015 &   0.017 &   0.006 &   0.007 &   0.029 &   0.011 &   0.006 &   0.006 &   0.008 \\
-0.3 & 0.8 &   0.015 &   0.044 &   0.030 &   0.015 &   0.013 &  -0.005 &   0.006 &   0.029 &   0.011 &   0.006 &   0.006 &   0.008 \\
 0.0 & 0.8 &   0.010 &   0.048 &   0.024 &   0.015 &   0.015 &  -0.013 &   0.006 &   0.029 &   0.010 &   0.006 &   0.006 &   0.009 \\
 0.3 & 0.8 &   0.011 &   0.044 &   0.017 &   0.016 &   0.017 &  -0.022 &   0.006 &   0.029 &   0.010 &   0.006 &   0.006 &   0.012 \\
 0.6 & 0.8 &   0.020 &   0.046 &   0.012 &   0.015 &   0.030 &  -0.049 &   0.007 &   0.028 &   0.010 &   0.006 &   0.006 &   0.036 \\
 1.2 & 0.8 &  -0.119 &   0.059 &   0.011 &   0.016 &   0.016 &  -0.118 &   0.021 &   0.029 &   0.010 &   0.006 &   0.006 &   0.022 \\
 1.8 & 0.8 &  -0.736 &   0.080 &   0.023 &   0.015 &   0.015 &  -0.732 & \largemse{  0.559} &   0.033 &   0.010 &   0.006 &   0.006 & \largemse{  0.557} \\
 2.2 & 0.8 &  -1.160 &   0.111 &   0.056 &   0.016 &   0.016 &  -1.157 & \largemse{  1.358} &   0.039 &   0.013 &   0.006 &   0.006 & \largemse{  1.354} \\
\bottomrule
\end{tabular}
\begin{tablenotes}
\footnotesize
\item Notes: Monte Carlo results for $\ARFIMA(0,d,1)$ processes with $n=500$, $m=56$, 10,000 replications.
    Shaded cells indicate $\text{MSE} > 0.05$.
\item LW = Local Whittle, V = Velasco (Kolmogorov), HC = Hurvich--Chen, ELW = Exact Local Whittle, 2ELW = Two-step ELW, LWLFC = Local Whittle robust to low-frequency contamination.
\end{tablenotes}
\end{threeparttable}
\end{table}

\begin{table}[tp]
\centering
\begin{threeparttable}
\caption{Estimator Comparison under ARMA(1,1) Short-Run Dynamics}
\label{tab:mc_comp_arma11}
\scriptsize
\begin{tabular}{ccc|rrrrrr|rrrrrr}
\toprule
&  &  & \multicolumn{6}{c|}{Bias} & \multicolumn{6}{c}{MSE} \\
\cmidrule(lr){4-9} \cmidrule(lr){10-15}
$d$ & $\phi$ & $\theta$ & LW & V & HC & ELW & 2ELW & LWLFC & LW & V & HC & ELW & 2ELW & LWLFC \\
\midrule
-2.2 & 0.5 &  0.5 &   1.209 &   0.210 &   0.842 &   0.117 &   1.044 &   1.201 & \largemse{  1.544} & \largemse{  0.072} & \largemse{  0.777} &   0.020 & \largemse{  1.151} & \largemse{  1.442} \\
-1.8 & 0.5 &  0.5 &   0.760 &   0.196 &   0.472 &   0.117 &   0.671 &   0.801 & \largemse{  0.640} & \largemse{  0.066} & \largemse{  0.257} &   0.020 & \largemse{  0.486} & \largemse{  0.641} \\
-1.2 & 0.5 &  0.5 &   0.261 &   0.179 &   0.193 &   0.115 &   0.268 &   0.236 & \largemse{  0.085} & \largemse{  0.059} &   0.047 &   0.019 & \largemse{  0.083} & \largemse{  0.059} \\
-0.6 & 0.5 &  0.5 &   0.125 &   0.171 &   0.150 &   0.117 &   0.116 &   0.116 &   0.022 & \largemse{  0.057} &   0.032 &   0.020 &   0.020 &   0.021 \\
-0.3 & 0.5 &  0.5 &   0.113 &   0.164 &   0.138 &   0.116 &   0.115 &   0.105 &   0.019 & \largemse{  0.054} &   0.029 &   0.020 &   0.020 &   0.018 \\
 0.0 & 0.5 &  0.5 &   0.110 &   0.167 &   0.133 &   0.116 &   0.117 &   0.101 &   0.018 & \largemse{  0.055} &   0.028 &   0.020 &   0.020 &   0.018 \\
 0.3 & 0.5 &  0.5 &   0.110 &   0.163 &   0.126 &   0.116 &   0.118 &   0.099 &   0.018 & \largemse{  0.054} &   0.026 &   0.020 &   0.021 &   0.018 \\
 0.6 & 0.5 &  0.5 &   0.116 &   0.169 &   0.122 &   0.116 &   0.119 &   0.084 &   0.020 & \largemse{  0.055} &   0.025 &   0.019 &   0.020 &   0.036 \\
 1.2 & 0.5 &  0.5 &  -0.092 &   0.174 &   0.120 &   0.116 &   0.116 &  -0.091 &   0.020 & \largemse{  0.057} &   0.025 &   0.020 &   0.020 &   0.020 \\
 1.8 & 0.5 &  0.5 &  -0.735 &   0.199 &   0.130 &   0.115 &   0.115 &  -0.732 & \largemse{  0.559} & \largemse{  0.067} &   0.027 &   0.019 &   0.019 & \largemse{  0.557} \\
 2.2 & 0.5 &  0.5 &  -1.158 &   0.229 &   0.161 &   0.116 &   0.116 &  -1.155 & \largemse{  1.355} & \largemse{  0.079} &   0.035 &   0.020 &   0.020 & \largemse{  1.351} \\
\midrule
-2.2 & 0.8 &  0.5 &   1.156 &   0.559 &   0.919 &   0.433 &   1.087 &   1.201 & \largemse{  1.404} & \largemse{  0.341} & \largemse{  0.890} & \largemse{  0.195} & \largemse{  1.220} & \largemse{  1.443} \\
-1.8 & 0.8 &  0.5 &   0.778 &   0.548 &   0.636 &   0.434 &   0.778 &   0.808 & \largemse{  0.642} & \largemse{  0.330} & \largemse{  0.421} & \largemse{  0.196} & \largemse{  0.626} & \largemse{  0.653} \\
-1.2 & 0.8 &  0.5 &   0.482 &   0.533 &   0.500 &   0.434 &   0.466 &   0.475 & \largemse{  0.241} & \largemse{  0.314} & \largemse{  0.260} & \largemse{  0.196} & \largemse{  0.223} & \largemse{  0.234} \\
-0.6 & 0.8 &  0.5 &   0.432 &   0.523 &   0.477 &   0.434 &   0.433 &   0.431 & \largemse{  0.194} & \largemse{  0.304} & \largemse{  0.239} & \largemse{  0.196} & \largemse{  0.195} & \largemse{  0.194} \\
-0.3 & 0.8 &  0.5 &   0.427 &   0.523 &   0.470 &   0.434 &   0.434 &   0.426 & \largemse{  0.189} & \largemse{  0.303} & \largemse{  0.233} & \largemse{  0.196} & \largemse{  0.196} & \largemse{  0.189} \\
 0.0 & 0.8 &  0.5 &   0.423 &   0.519 &   0.463 &   0.434 &   0.437 &   0.423 & \largemse{  0.186} & \largemse{  0.299} & \largemse{  0.226} & \largemse{  0.195} & \largemse{  0.200} & \largemse{  0.186} \\
 0.3 & 0.8 &  0.5 &   0.421 &   0.519 &   0.458 &   0.432 &   0.434 &   0.420 & \largemse{  0.184} & \largemse{  0.298} & \largemse{  0.222} & \largemse{  0.195} & \largemse{  0.196} & \largemse{  0.185} \\
 0.6 & 0.8 &  0.5 &   0.412 &   0.521 &   0.455 &   0.434 &   0.434 &   0.409 & \largemse{  0.177} & \largemse{  0.300} & \largemse{  0.219} & \largemse{  0.196} & \largemse{  0.196} & \largemse{  0.181} \\
 1.2 & 0.8 &  0.5 &  -0.037 &   0.529 &   0.453 &   0.433 &   0.433 &  -0.036 &   0.032 & \largemse{  0.309} & \largemse{  0.217} & \largemse{  0.195} & \largemse{  0.195} &   0.033 \\
 1.8 & 0.8 &  0.5 &  -0.732 &   0.550 &   0.460 &   0.434 &   0.434 &  -0.728 & \largemse{  0.558} & \largemse{  0.331} & \largemse{  0.223} & \largemse{  0.196} & \largemse{  0.196} & \largemse{  0.556} \\
 2.2 & 0.8 &  0.5 &  -1.160 &   0.569 &   0.481 &   0.434 &   0.433 &  -1.157 & \largemse{  1.358} & \largemse{  0.351} & \largemse{  0.241} & \largemse{  0.196} & \largemse{  0.195} & \largemse{  1.355} \\
\midrule
-2.2 & 0.9 &  0.5 &   1.182 &   0.803 &   1.039 &   0.681 &   1.163 &   1.204 & \largemse{  1.448} & \largemse{  0.675} & \largemse{  1.113} & \largemse{  0.472} & \largemse{  1.379} & \largemse{  1.450} \\
-1.8 & 0.9 &  0.5 &   0.870 &   0.794 &   0.823 &   0.684 &   0.875 &   0.851 & \largemse{  0.777} & \largemse{  0.661} & \largemse{  0.689} & \largemse{  0.476} & \largemse{  0.779} & \largemse{  0.729} \\
-1.2 & 0.9 &  0.5 &   0.701 &   0.783 &   0.741 &   0.684 &   0.688 &   0.700 & \largemse{  0.499} & \largemse{  0.644} & \largemse{  0.561} & \largemse{  0.476} & \largemse{  0.481} & \largemse{  0.498} \\
-0.6 & 0.9 &  0.5 &   0.676 &   0.774 &   0.722 &   0.682 &   0.682 &   0.675 & \largemse{  0.464} & \largemse{  0.629} & \largemse{  0.533} & \largemse{  0.473} & \largemse{  0.473} & \largemse{  0.464} \\
-0.3 & 0.9 &  0.5 &   0.672 &   0.771 &   0.717 &   0.683 &   0.686 &   0.672 & \largemse{  0.460} & \largemse{  0.625} & \largemse{  0.525} & \largemse{  0.474} & \largemse{  0.479} & \largemse{  0.460} \\
 0.0 & 0.9 &  0.5 &   0.670 &   0.776 &   0.714 &   0.683 &   0.686 &   0.670 & \largemse{  0.457} & \largemse{  0.632} & \largemse{  0.521} & \largemse{  0.474} & \largemse{  0.478} & \largemse{  0.456} \\
 0.3 & 0.9 &  0.5 &   0.665 &   0.773 &   0.709 &   0.684 &   0.685 &   0.662 & \largemse{  0.449} & \largemse{  0.627} & \largemse{  0.515} & \largemse{  0.475} & \largemse{  0.477} & \largemse{  0.450} \\
 0.6 & 0.9 &  0.5 &   0.611 &   0.775 &   0.705 &   0.684 &   0.684 &   0.612 & \largemse{  0.383} & \largemse{  0.630} & \largemse{  0.509} & \largemse{  0.475} & \largemse{  0.476} & \largemse{  0.384} \\
 1.2 & 0.9 &  0.5 &  -0.022 &   0.786 &   0.704 &   0.684 &   0.684 &  -0.019 &   0.047 & \largemse{  0.646} & \largemse{  0.508} & \largemse{  0.476} & \largemse{  0.475} &   0.048 \\
 1.8 & 0.9 &  0.5 &  -0.732 &   0.802 &   0.709 &   0.683 &   0.682 &  -0.728 & \largemse{  0.559} & \largemse{  0.670} & \largemse{  0.515} & \largemse{  0.475} & \largemse{  0.473} & \largemse{  0.557} \\
 2.2 & 0.9 &  0.5 &  -1.157 &   0.821 &   0.722 &   0.683 &   0.682 &  -1.154 & \largemse{  1.354} & \largemse{  0.702} & \largemse{  0.531} & \largemse{  0.474} & \largemse{  0.472} & \largemse{  1.351} \\
\bottomrule
\end{tabular}
\begin{tablenotes}
\footnotesize
\item Notes: Monte Carlo results for $\ARFIMA(1,d,1)$ processes with $n=500$, $m=56$, 10,000 replications.
    Shaded cells indicate $\text{MSE} > 0.05$.
\item LW = Local Whittle, V = Velasco (Kolmogorov), HC = Hurvich--Chen, ELW = Exact Local Whittle, 2ELW = Two-step ELW, LWLFC = Local Whittle robust to low-frequency contamination.
\end{tablenotes}
\end{threeparttable}
\end{table}

\begin{table}[tp]
\centering
\begin{threeparttable}
\caption{Estimator Comparison under Heavy-Tailed and GARCH Innovations}
\label{tab:mc_comp_heavy}
\scriptsize
\begin{tabular}{c|rrrrrr|rrrrrr}
\toprule
& \multicolumn{6}{c|}{Bias} & \multicolumn{6}{c}{MSE} \\
\cmidrule(lr){2-7} \cmidrule(lr){8-13}
$d$ & LW & V & HC & ELW & 2ELW & LWLFC & LW & V & HC & ELW & 2ELW & LWLFC \\
\midrule
\multicolumn{13}{l}{\textit{Student-$t(5)$ innovations}} \\
-2.2 &   1.480 &   0.072 &   0.971 &  -0.002 &   1.176 &   1.258 & \largemse{  2.291} &   0.032 & \largemse{  1.029} &   0.006 & \largemse{  1.457} & \largemse{  1.606} \\
-1.8 &   0.968 &   0.055 &   0.533 &  -0.004 &   0.738 &   0.824 & \largemse{  1.026} &   0.029 & \largemse{  0.343} &   0.006 & \largemse{  0.596} & \largemse{  0.687} \\
-1.2 &   0.292 &   0.039 &   0.108 &  -0.001 &   0.259 &   0.228 & \largemse{  0.124} &   0.028 &   0.024 &   0.006 & \largemse{  0.085} & \largemse{  0.057} \\
-0.6 &   0.017 &   0.027 &   0.022 &  -0.003 &   0.001 &  -0.010 &   0.007 &   0.027 &   0.010 &   0.006 &   0.006 &   0.008 \\
-0.3 &  -0.002 &   0.025 &   0.012 &  -0.002 &  -0.004 &  -0.024 &   0.006 &   0.027 &   0.010 &   0.006 &   0.006 &   0.008 \\
 0.0 &  -0.007 &   0.024 &   0.004 &  -0.002 &  -0.002 &  -0.032 &   0.006 &   0.027 &   0.010 &   0.006 &   0.006 &   0.010 \\
 0.3 &  -0.009 &   0.021 &  -0.004 &  -0.004 &  -0.003 &  -0.047 &   0.006 &   0.027 &   0.010 &   0.006 &   0.006 &   0.014 \\
 0.6 &   0.003 &   0.028 &  -0.005 &  -0.002 &   0.013 &  -0.078 &   0.007 &   0.027 &   0.010 &   0.006 &   0.006 &   0.045 \\
 1.2 &  -0.125 &   0.038 &  -0.008 &  -0.003 &  -0.003 &  -0.124 &   0.022 &   0.028 &   0.009 &   0.006 &   0.006 &   0.024 \\
 1.8 &  -0.736 &   0.065 &   0.005 &  -0.002 &  -0.002 &  -0.732 & \largemse{  0.559} &   0.031 &   0.009 &   0.006 &   0.006 & \largemse{  0.557} \\
 2.2 &  -1.161 &   0.091 &   0.040 &  -0.002 &  -0.002 &  -1.158 & \largemse{  1.359} &   0.035 &   0.011 &   0.006 &   0.006 & \largemse{  1.355} \\
\midrule
\multicolumn{13}{l}{\textit{GARCH(1,1) innovations}} \\
-2.2 &   1.489 &   0.072 &   0.984 &  -0.003 &   1.187 &   1.260 & \largemse{  2.316} &   0.033 & \largemse{  1.052} &   0.007 & \largemse{  1.484} & \largemse{  1.609} \\
-1.8 &   0.986 &   0.056 &   0.545 &  -0.004 &   0.750 &   0.825 & \largemse{  1.057} &   0.031 & \largemse{  0.353} &   0.007 & \largemse{  0.613} & \largemse{  0.689} \\
-1.2 &   0.314 &   0.036 &   0.111 &  -0.003 &   0.270 &   0.229 & \largemse{  0.136} &   0.030 &   0.025 &   0.007 & \largemse{  0.090} & \largemse{  0.057} \\
-0.6 &   0.018 &   0.026 &   0.022 &  -0.003 &   0.002 &  -0.009 &   0.008 &   0.028 &   0.011 &   0.007 &   0.006 &   0.009 \\
-0.3 &  -0.003 &   0.025 &   0.012 &  -0.003 &  -0.005 &  -0.025 &   0.007 &   0.029 &   0.011 &   0.007 &   0.007 &   0.010 \\
 0.0 &  -0.007 &   0.022 &   0.004 &  -0.003 &  -0.002 &  -0.034 &   0.007 &   0.029 &   0.011 &   0.007 &   0.007 &   0.011 \\
 0.3 &  -0.008 &   0.018 &  -0.004 &  -0.003 &  -0.002 &  -0.048 &   0.007 &   0.029 &   0.011 &   0.007 &   0.007 &   0.016 \\
 0.6 &   0.003 &   0.027 &  -0.006 &  -0.002 &   0.012 &  -0.085 &   0.007 &   0.030 &   0.011 &   0.007 &   0.007 & \largemse{  0.051} \\
 1.2 &  -0.124 &   0.039 &  -0.008 &  -0.004 &  -0.004 &  -0.124 &   0.022 &   0.029 &   0.011 &   0.007 &   0.007 &   0.026 \\
 1.8 &  -0.734 &   0.066 &   0.006 &  -0.002 &  -0.002 &  -0.730 & \largemse{  0.558} &   0.032 &   0.010 &   0.007 &   0.007 & \largemse{  0.555} \\
 2.2 &  -1.161 &   0.090 &   0.039 &  -0.002 &  -0.002 &  -1.158 & \largemse{  1.360} &   0.036 &   0.012 &   0.007 &   0.007 & \largemse{  1.356} \\
\bottomrule
\end{tabular}
\begin{tablenotes}
\footnotesize
\item Notes: Monte Carlo results for $\ARFIMA(0,d,0)$ processes with $n=500$, $m=56$, 10,000 replications,
    and mean-zero, unit-variance innovations $\varepsilon_t$.
    Student-$t(5)$: $\varepsilon_t = \eta_t / \sqrt{\nf{5}{3}}$ with
    $\eta_t \iidsim t(5)$ (kurtosis 9).
    GARCH(1,1): $\varepsilon_t = \sigma_t \eta_t$ with $\eta_t \iidsim \Normal(0,1)$ and
    $\sigma_t^2 = \omega + \alpha \varepsilon_{t-1}^2 + \beta \sigma_{t-1}^2$,
    $(\omega, \alpha, \beta) = (0.1, 0.1, 0.8)$,
    implying unit unconditional variance and kurtosis $\approx 3.35$.
    Shaded cells indicate $\text{MSE} > 0.05$.
\item LW = Local Whittle, V = Velasco (Kolmogorov), HC = Hurvich--Chen, ELW = Exact Local Whittle, 2ELW = Two-step ELW, LWLFC = Local Whittle robust to low-frequency contamination.
\end{tablenotes}
\end{threeparttable}
\end{table}
}

\clearpage
\section{Comparison of Velasco Tapers}
\label{sec:appendix:tapers}

Table~\ref{tab:mc:lw_v_all} compares the finite-sample performance of
the three tapers considered by \cite{velasco-1999}, using the same
Monte Carlo design as
Table~\ref{tab:mc:replications:lw}.\footnote{We note that the cosine bell
  taper is labeled $p=3$, but this denotes the frequency subsampling
  interval rather than the order of the taper, which is one.}
The results reflect the fundamental trade-off between
bias reduction and variance inflation due to tapering.
Within the range $d \in [-0.7, 0.7]$, all three tapers perform
well.
The Bartlett taper ($p=2$) achieves the lowest variance
with standard deviations around 0.120.

We can see the advantages of higher-order tapering in the nonstationary range.
The order-three Kolmogorov taper retains low bias throughout
$d \in [1.3, 2.3]$ and improves on the Bartlett taper at the top of this
range, where its bias at $d = 2.3$ is $0.102$ against $-0.177$ for Bartlett,
though at the cost of increased variance, with standard deviations around
$0.16$, roughly 40\% above the theoretical value of
$\sqrt{\nf{3.01}{4m}} = 0.116$.
The cosine bell tracks the Kolmogorov taper closely while both remain in
range, but its order-one validity becomes apparent by $d = 2.3$, where its
bias exceeds that of the Bartlett and Kolmogorov tapers.
At still more extreme values such as $d = 3.5$, all three tapers break down
well outside their valid ranges.

\begin{table}[tp]
\centering
\begin{threeparttable}
\caption{Comparison of Velasco (1999) Tapers}
\label{tab:mc:lw_v_all}
\begin{tabular}{r@{\hspace{1em}}rrr@{\hspace{1em}}rrr@{\hspace{1em}}rrr}
\toprule
& \multicolumn{3}{c}{Bartlett ($p=2$)} & \multicolumn{3}{c}{Cosine ($p=3$)} & \multicolumn{3}{c}{Kolmogorov ($p=3$)} \\
\cmidrule(lr){2-4} \cmidrule(lr){5-7} \cmidrule(lr){8-10}
$d$ & Bias & S.D. & MSE & Bias & S.D. & MSE & Bias & S.D. & MSE \\
\midrule
$-3.5$ & $ 1.6231$ & $ 0.3326$ & $\largemse{  2.745}$ & $ 0.1254$ & $ 0.1644$ & $  0.043$ & $ 0.2207$ & $ 0.1796$ & $\largemse{  0.081}$ \\
$-2.3$ & $ 0.2173$ & $ 0.1734$ & $\largemse{  0.077}$ & $ 0.0698$ & $ 0.1644$ & $  0.032$ & $ 0.0757$ & $ 0.1648$ & $  0.033$ \\
$-1.7$ & $ 0.0279$ & $ 0.1211$ & $  0.015$ & $ 0.0515$ & $ 0.1643$ & $  0.030$ & $ 0.0542$ & $ 0.1649$ & $  0.030$ \\
$-1.3$ & $ 0.0090$ & $ 0.1203$ & $  0.015$ & $ 0.0419$ & $ 0.1642$ & $  0.029$ & $ 0.0431$ & $ 0.1648$ & $  0.029$ \\
$-0.7$ & $-0.0048$ & $ 0.1200$ & $  0.014$ & $ 0.0315$ & $ 0.1640$ & $  0.028$ & $ 0.0313$ & $ 0.1645$ & $  0.028$ \\
$-0.3$ & $-0.0097$ & $ 0.1201$ & $  0.015$ & $ 0.0274$ & $ 0.1636$ & $  0.028$ & $ 0.0268$ & $ 0.1640$ & $  0.028$ \\
$ 0.0$ & $-0.0111$ & $ 0.1201$ & $  0.015$ & $ 0.0258$ & $ 0.1632$ & $  0.027$ & $ 0.0255$ & $ 0.1635$ & $  0.027$ \\
$ 0.3$ & $-0.0103$ & $ 0.1201$ & $  0.015$ & $ 0.0257$ & $ 0.1629$ & $  0.027$ & $ 0.0259$ & $ 0.1631$ & $  0.027$ \\
$ 0.7$ & $-0.0051$ & $ 0.1203$ & $  0.014$ & $ 0.0281$ & $ 0.1625$ & $  0.027$ & $ 0.0296$ & $ 0.1626$ & $  0.027$ \\
$ 1.3$ & $ 0.0170$ & $ 0.1222$ & $  0.015$ & $ 0.0410$ & $ 0.1628$ & $  0.028$ & $ 0.0432$ & $ 0.1622$ & $  0.028$ \\
$ 1.7$ & $ 0.0485$ & $ 0.1258$ & $  0.018$ & $ 0.0722$ & $ 0.1654$ & $  0.033$ & $ 0.0597$ & $ 0.1620$ & $  0.030$ \\
$ 2.3$ & $-0.1774$ & $ 0.1410$ & $\largemse{  0.051}$ & $ 0.2316$ & $ 0.2010$ & $\largemse{  0.094}$ & $ 0.1020$ & $ 0.1636$ & $  0.037$ \\
$ 3.5$ & $-1.4529$ & $ 0.1378$ & $\largemse{  2.130}$ & $-0.3994$ & $ 0.1041$ & $\largemse{  0.170}$ & $-0.4160$ & $ 0.1835$ & $\largemse{  0.207}$ \\
\bottomrule
\end{tabular}
\begin{tablenotes}
\footnotesize
\item Notes: Comparison of three Velasco tapers, $n = 500$ observations from $\ARFIMA(0,d,0)$, $m = \lfloor n^{0.65} \rfloor = 56$ frequencies, 10,000 replications. Shaded cells indicate $\text{MSE} > 0.05$.
\end{tablenotes}
\end{threeparttable}
\end{table}

\end{document}